\documentclass[12pt,%twocolumn
]{revtex4-1}
\usepackage{amssymb,amsmath}
\usepackage{subfig}
\usepackage{graphicx}
\usepackage{color}
\usepackage{hyperref}
\usepackage{listings}
\usepackage[utf8]{inputenc}
\usepackage{dsfont}
\usepackage{braket}
\usepackage[normalem]{ulem}
\hypersetup{
    bookmarks=true,         % show bookmarks bar?
    unicode=false,          % non-Latin characters in Acrobat’s bookmarks
    pdftoolbar=true,        % show Acrobat’s toolbar?
    pdfmenubar=true,        % show Acrobat’s menu?
    pdffitwindow=false,     % window fit to page when opened
    pdfstartview={FitH},    % fits the width of the page to the window
    pdfauthor={Kevin Multani},     % author
    colorlinks=true,       % false: boxed links; true: colored links
    linkcolor=blue,          % color of internal links
    citecolor=red,        % color of links to bibliography
}
\graphicspath{{figures/}}

\begin{document}

\definecolor{dkgreen}{rgb}{0,0.6,0}
\definecolor{gray}{rgb}{0.5,0.5,0.5}
\definecolor{mauve}{rgb}{0.58,0,0.82}

\def\H{{\mathbf H}}
\def\One{{\boldsymbol\sigma}_0}
\def\be{\begin{equation}}
\def\ee{\end{equation}}
\def\bea{\begin{eqnarray}}
\def\eea{\end{eqnarray}}
\def\nnb{\nonumber}
\def\<{\langle}
\def\>{\rangle}

%\lstset{frame=tb,
 % 	language=Matlab,
 % 	aboveskip=3mm,
 % 	belowskip=3mm,
 % 	showstringspaces=false,
 % 	columns=flexible,
 % 	basicstyle={\small\ttfamily},
 % 	numbers=none,
 % 	numberstyle=\tiny\color{gray},
 %	keywordstyle=\color{blue},
%	commentstyle=\color{dkgreen},
 % 	stringstyle=\color{mauve},
  %	breaklines=true,
  %	breakatwhitespace=true
  %	tabsize=3
%}

\title{Spin current generation and control in carbon nanotubes by combining rotation and magnetic field  }

\author{Márcio M. Cunha}
\affiliation{Departamento de Física, CCET,  Universidade Federal 
do Maranhão, 65085-580 , São Luís, MA, Brazil}

\author{Jonas R. F. Lima
}
\affiliation{Institute of Nanotechnology, Karlsruhe Institute of Technology, D-76021 Karlsruhe, Germany}
\affiliation{Departamento de F\'{\i}sica, Universidade Federal Rural de Pernambuco, 
52171-900, Recife, PE, Brazil}

\author{Fernando Moraes}
\affiliation{Departamento de F\'{\i}sica, Universidade Federal Rural de Pernambuco, 
52171-900, Recife, PE, Brazil}

\author{S\'ebastien Fumeron and Bertrand Berche}
\affiliation{Laboratoire de Physique et Chimie Th\'eoriques,  UMR Universit\'e de Lorraine - CNRS 7019,
 54506 Vand\oe uvre les Nancy, France}

\date{\today}

\begin{abstract}
We study the quantum dynamics of ballistic electrons in rotating carbon nanotubes in the presence of a uniform magnetic field.  When the field is parallel to the nanotube axis, the  rotation-induced electric field  brings about the spin-orbit interaction which, together with the kinetic, inertial, and Zeeman terms, compose the Schr\"odinger-Pauli Hamiltonian of the system.  Full diagonalization of this  Hamiltonian yields the eigenstates and eigenenergies leading to the calculation of the charge and spin currents. Our main result is the demonstration that, by suitably combining the applied magnetic field intensity and rotation speed, one can tune one of the currents to zero while keeping the other one finite, giving rise to a spin current generator.  

 %We are interested in the effect of both electromagnetic and inertial effects on the quantum dynamics of a charged particle. In the formalism, there are analogies on the way in which inertial effects are introduced and the minimal coupling to an electromagnetic field. In this paper, we solve the Schr\"odinger-Pauli equation for several cases taking into account rotation and the action of a magnetic field. We solve the eigenvalue problem and contemplate the resulting charge and spin currents. We observe that, for a specific combination of  the applied magnetic field, the rotation speed and injection momentum, specified by an algebraic  equation, the charge current can be eliminated leaving a non-null spin current.

%{\bf R\'esum\'e :} Nous nous int\'eressons aux consequences combin\'ees de la pr\'esence d'un champ \'electromagn\'etique et des effets inertiels sur la dynamique quantique d'une particule charg\'ee. Dans le formalisme, il y a des analogies entre la mani\`ere dont sont introduits les effets inertiels et le couplage minimal \`a un champ \'electromagn\'etique. Dans cet article, on r\'esoud l'\'equation de Pauli-Schr\"odinger dans diff\'erents cas en prenant en compte la rotation et la pr\'esence d'un champ magn\'etique. On r\'esoud le probl\`eme aux valeurs propres et on d\'etermine les courants de charge et de spin r\'esultants.

\end{abstract}

\maketitle

\section{Introduction}

Carbon materials are ubiquitous, ranging from living organisms to contemporary high technology devices. Pure carbon forms have fascinated humanity for ages, starting with diamonds and culminating in the last decades with nanoscale structures like graphene, fullerenes and nanotubes. These nanostructures have attracted enormous attention lately due to their unusual physical properties which, not only provide a thrilling laboratory for fundamental physics, but also lead to important technological applications. In particular, carbon nanotubes applications range from water treatment, to composite materials with special thermal or mechanical properties, and to electronics, among others.  For a recent review of the latter, see \cite{cao2019review}. Nanomechanical applications have also been contemplated. For example
nanomotors made of Carbon nanotubes and diamond needles have been proposed, see e.g. 
\cite{LI2019260}. The rotation frequency in these devices may reach values as large as $1-100$~GHz. Also, nano-turbines composed of carbon nanotubes and graphene nanoblades have been designed \cite{li2014rotation}.  Other aspects and applications involving rotation in nanosystems  have also been investigated. Ref. \cite{narendar2011nonlocal}, for instance, deals with wave propagation in a rotating nanotube and in \cite{belhadj2017free}, it was investigated the vibrational behaviour of a rotating shaft based single-walled carbon nanotube. Reference \cite{wang2017design} proposes  a nano screw pump by use of rotating helical nanowires  and Ref. \cite{tu2016rotating} presents the design of a water desalination device using rotating nanotubes.

Analogous to electronics, spintronics \cite{vzutic2004spintronics}, which is based on the spin degree of freedom of the electron instead of the charge, is driven by spin currents which may or may not be accompanied by  charge currents. Like its sister technology, spintronics is appearing as an important source of novel devices \cite{joshi2016spintronics}. It has been recently shown \cite{guimaraes2010carbon} that carbon nanotubes may be excellent spin current waveguides.  Additionally, as shown in Ref. \cite{kral2002laser}, carbon nanotubes can be lead to spin at GHz frequencies by circularly polarized light. Also, rotation and magnetic field have striking similarities (see for instance \cite{brandao2015inertial} and references therein). For example, rotation couples to spin leading to the celebrated Barnett (magnetization by rotation) and Einstein-de Haas (rotation by magnetization) effects. These facts motivated us to investigate the combined effects of electromagnetic fields and rotation on the electronic energy eigenvalues and on the generation and control of charge and spin currents in carbon nanotubes, which are known to be good ballistic electron conductors  \cite{poncharal2002roomACS,white1998carbon}. Since ballistic transport occurs in  high energy bands, it can be studied with the help of the Schr\"odinger equation (see, for instance,   section 8.1.1 on the Ref. \cite{dresselhaus1998physical}) , while the electronic properties  near the Fermi level are well described by the massless Dirac equation \cite{ando2000theory}. The latter has been used in previous works on rotating fullerenes \cite{lima2014effects,lima2015combined} and carbon nanotubes \cite{cunha2015spin} to study inertial effects on their low-energy excitations.

In this paper, we solve the Schrödinger-Pauli equation for a free electron confined to a rotating nanotube, taking into account the influence of both electromagnetic fields and inertial effects in the energy spectrum and generation of spin and charge currents.  While spin-rotation coupling, via a twisting phonon mode,  has been recently proposed \cite{hamada2015spin} as means of generating spin currents in nanotubes, we consider here a rigid nanotube under external rotation which may be caused by circularly polarized light \cite{kral2002laser}, for instance.   We study two different configurations for a nanotube rotating around its symmetry axis. In the first one, an external magnetic field parallel to the tube axis induces, in the rotating frame, a radial electric field  which, by its turn, switches on the spin-orbit coupling on the electrons.  Under these circumstances, injection of ballistic electrons in one of the extremities of the nanotube leads to both a spin and a charge current. We show that, by a suitable choice of magnetic field, rotation speed and injection momentum, the charge current can be brought down to zero, leaving a pure spin current in the system. On the other hand, the spin current can also be tuned to zero while the charge current is kept finite. In the second situation, the magnetic field is azimuthal,  inducing an axial electric field which does not lead to spin-orbit coupling. In this case, the $z$ component of the spin current is proportional to the corresponding component of the charge current. Therefore both are tuned  to zero simultaneously and consequently this field configuration is not interesting for current management.  This way, we will give a special attention to the axial case along the paper.

{The paper is organized in the following way: in section II, we derive the Schrödinger-Pauli equation for a rotating frame to accommodate the interactions involving the spin of the electron. In section III, we obtain the energy spectrum and the eigenfunctions for a particle in a rotating nanotube in the presence of an axial magnetic field. In section IV, we obtain the charge and spin current densities corresponding to  the same field configuration.} In section V we present our conclusions and in the Appendix, for the sake of completeness, we present  the results concerning an azimuthal magnetic field.

\section{Schrödinger-Pauli equation in a rotating frame}

Following the approach described in \cite{matsuo2011spin}, we will start from the equation of motion
\begin{equation}
\H\Psi=i\hbar \frac{\partial \Psi}{\partial t},
\label{hamiltoniano_geral}
\end{equation}
where $\Psi$ is the two-component spinor living in the Hilbert space  $\mathfrak{H}={\cal L}^2({\mathbb R}^3)\otimes{\mathbb C}^2$ where ${\cal L}^2({\mathbb R}^3)$ is the set of square-integrable complex functions over $\mathbb R^3$   and $\H$ is the generator of the dynamics, the Hamiltonian (bold characters denote $2\times 2$ matrices acting on the spinors in ${\mathbb C}^2$). The Hamiltonian $\H$ contains several contributions:
\begin{equation}
\H=\H_{\rm K} + \H_{\rm I}+ \H_{\rm Z}  + \H_{\rm SO}.
\end{equation}
This Hamiltonian describes the quantum behaviour of an electron of charge $q=-|e|$ and spin $\vec{\mathbf s}=\frac 12\hbar\vec{\boldsymbol{\sigma}}$.
In this work, we will consider that the electron is subject to move on the surface of a nanotube of radius $\rho=a$ oriented such as its symmetry axis coincides with the $z$-axis. The tube rotates around its symmetry axis at a constant angular velocity $\omega$.

Let us now define the terms in the above Hamiltonian. 
The  term $\H_{\rm K}$ corresponds to the kinetic energy (KE) plus diagonal terms for convenience, electrostatic energy $-|e|A_0$ and the da Costa potential \cite{da1981quantum}, which summarizes here to a constant term $-\hbar^2/(8ma^2)$, but the presence of which would introduce a $z-$varying potential if the tube had corrugations \cite{Santos_2016,Fumeron_2017,serafim2019position}. 
Altogether, this contribution to the Hamiltonian is proportional to identity in spin space and
is given by
\begin{equation}
\H_{\rm K}=\left(\frac{1}{2m}|\vec\Pi|^2  -|e|A_0-\frac{\hbar^2}{8ma^2}\right)\One,
\end{equation}
where $\vec{\Pi}=\vec{p}+|e|\vec{A}$ is the mechanical momentum, defined in terms of the canonical momentum $\vec p$ through minimal coupling, $\vec{A}$ is the vector potential, $A_0$ is the scalar potential and $\One$ is the $2\times 2$ identity matrix in spin space.
The second term, $\H_{\rm I}$, contains inertial effects, i.e., the coupling between  both the orbital degrees of freedom and  the spin with  rotation. We will consider here, as already mentioned, the case of a nanotube rotating around its symmetry axis, $\vec{\omega}=\omega\hat z$. Thus, in cylindrical coordinates $(\rho,\varphi;z)$:
\begin{equation}
\H_{\rm I}=-\vec\omega\cdot[(\vec{r} \times \vec{\Pi})\One + \vec{\mathbf s}\ \!  ].
\end{equation}
Note that the kinetic energy and the coupling of orbital degrees of freedom with rotation can be written in a canonical manner
\begin{equation}
\frac{1}{2m}|\vec\Pi|^2 -\vec\omega\cdot(\vec{r} \times \vec{\Pi})=
\frac 1{2m}(\vec\Pi-m\vec\omega\times\vec r)^2-\frac 12m(\vec \omega\times\vec r)^2.
\label{Eq-RotationGauge}
\end{equation}
The term $\H_{\rm Z}$ corresponds to the Zeeman interaction which couples the electron spin to the magnetic field:
\begin{equation}
\H_{\rm Z} = \mu_B \vec{\boldsymbol{\sigma}}\cdot \vec{B},
\end{equation}
where the Bohr magneton is $\mu_B=\frac{|e|\hbar}{2m}$,  $\vec{B}$ is the magnetic field and $\vec{\boldsymbol{\sigma}}=(\boldsymbol{\sigma}_x, \boldsymbol{\sigma}_y, \boldsymbol{\sigma}_z)$ the vector of Pauli matrices. 
Eventually, the spin-orbit interaction is given after proper symmetrization by 
\begin{equation}
\H_{\rm SO}=-\frac 12\kappa \vec{\boldsymbol{\sigma}}.\left(\vec{\Pi} \times \vec{E}'- \vec{E}' \times \vec{\Pi} \right),\label{H-rotation}
\end{equation}
with
\begin{equation}
\kappa= \frac{|e|\hbar}{4m^2 c^2}.    
\end{equation}
$\vec E'$ is the electric field  in the rotating frame, given in terms of  $\vec{E}$, the applied electric field in the inertial laboratory frame by
\begin{equation}
\vec{E}'=\vec{E}+ (\vec{\omega} \times \vec{r}) \times \vec{B}.\label{eq-Eprime}
\end{equation}
The term $(\vec{\omega} \times \vec{r}) \times \vec{B}$  is the electric field due to rotation, our main interest in this study. Thus, we will consider $\vec{E}=\vec{0}$. In the present symmetry,
$\vec\omega\times\vec r=\omega a\hat\varphi$ and,
if we only consider uniform magnetic fields, one has
 $(\vec p+|e|\vec A)\times[(\vec\omega\times \vec r)\times\vec B]=-[(\vec\omega\times \vec r)\times\vec B]\times (\vec p+|e|\vec A)$ and it follows that
\begin{equation}
  \H_{\rm SO}=-\kappa \vec{\boldsymbol{\sigma}}\cdot [(\vec p+|e|\vec A)\times (\omega a\hat\varphi\times\vec B)].
 \end{equation}
Note that the spin-orbit term here follows from the fact that, due to rotation, the electron experiences an associated electric field although only a magnetic field is applied  in the rest frame. 

We omit other contributions coming from the non-relativistic limit of Dirac equation like the Darwin term or the corrections to kinetic energy.
The complete Hamiltonian finally reads as
\bea
\H&=&\left(\frac{1}{2m}|\vec p+|e|\vec A|^2  -|e|A_0-\frac{\hbar^2}{8ma^2}\right)\One
-\vec\omega\cdot[\vec{r} \times (\vec p+|e|\vec A)  ]\One\nnb
\\
&&-\frac 12\hbar \vec\omega\cdot  \vec{\boldsymbol{\sigma}}+ \frac{|e|\hbar}{2m} \vec{\boldsymbol{\sigma}}\cdot \vec{B}
-\frac{|e|\hbar}{4m^2 c^2} \vec{\boldsymbol{\sigma}}\cdot [(\vec p+|e|\vec A)\times (\omega a\hat\varphi\times\vec B)],\label{eq-Htot}
\eea
where we have written  separately on purpose the purely orbital part from the part which explicitly  involves spin.

\section{Energy spectrum} 
In this section, we diagonalize the Schr\"odinger-Pauli Hamiltonian (\ref{eq-Htot}) for the case of an axial magnetic field. While the axial field includes  spin-orbit coupling due to the induced radial electric field caused by the rotating frame, in the azimuthal case this interaction does not exist, for the induced electric field is absent (see Eq. (\ref{eq-Eprime})).
For this reason, we will focus on the axial case here. We deal with the azimuthal case in the Appendix.  %Thus, we will be able to analyse the similarities and differences between the effects of rotation and those of the magnetic field as well as their combined effect for the  dynamics of the particle. In order to obtain the energy spectrum for the particle in the presence of a magnetic field and of  rotation, we  consider the Pauli-Schr\"odinger Hamiltonian, equation (\ref{hamiltoniano_geral}) 
In what follows we will be using  the cylindrical coordinates $\rho$, $ \varphi$ and $z$, with
\be
\vec p=-i\hbar (\rho^{-1}\partial_\varphi) \hat\varphi-i\hbar (\partial_z)\hat z.
\ee

%\subsection{Axial magnetic field} 
In the following, we specialize to the case of an axial magnetic field where  $\vec{B}
=B\hat z$ is uniform in which case we will use the symmetric gauge $\vec{A}=\frac 12\vec{B}\times\vec{r}=\frac 12Ba\hat\varphi$.

\subsection{A comment on the choice of gauge }

We  first discuss the case of the purely orbital motion i.e., cancel all terms involving $\vec{\boldsymbol\sigma}$ in $\H$ in (\ref{eq-Htot}). Then, an interesting property arises when we look at the formulation of Eq.~(\ref{Eq-RotationGauge}). The last term in the R.H.S (interpreted as a gauge symmetry breaking (GSB) term, see e.g. \cite{0295-5075-83-4-47005} and \cite{0295-5075-97-6-67007} for a similar discussion on the role of such a term in the case of spin-orbit interactions) is constant in our case and the first term becomes $\frac 1{2m}\bigl[\vec p+(\frac{|e|Ba}2-m\omega a)\hat\varphi\bigr]^2$. It immediately follows that rotation kills the effect of the magnetic field when $|e|B/2=m\omega$, hence when the angular frequency equals the Larmor frequency (half the cyclotron frequency)
$\omega=\omega_L=\frac 12\omega_c=|e|B/2m$.
An interesting comment here concerns the choice of gauge. Although gauge invariance guarantees that the above result remains correct with another gauge choice (see e.g. Ref.~\cite{doi:10.1119/1.4955153} for extended discussion),  we see that   with the Landau gauge $\vec A=Bx\hat y$, for instance, nothing special seems to happen in Eq.~(\ref{Eq-RotationGauge}) at $\omega_L$, which means that the property mentioned above is hidden in that case.
Let us mention also that motion on a more general cylindrically symmetric system, like a cylinder with bumps or hollows \cite{Santos_2016}, would not exhibit the property that  magnetic field effects  may be compensated by simple rotation, due to the presence of the GSB term which, then, would depend on $z$ and would then alter the form of the wavefunctions and energies.

\subsection{Eigenenergies}

Specializing to the cylindrical coordinates and $\rho = a$, the various terms of the Hamiltonian can be explicitly written as
\begin{eqnarray}
\H_{\rm K}&=&\frac 1{2m}\Bigl[\bigl(
-i\hbar a^{-1}\partial_\varphi+\frac 12|e|Ba\bigr)^2+(-i\hbar\partial_z)^2-\frac 14\hbar^2a^{-2}
\Bigr]\One,\\
\H_{\rm I}&=&-\omega a\Bigl[-i\hbar a^{-1}\partial_\varphi
+\frac 12|e|Ba\Bigr]\One-\frac 12\hbar \omega\boldsymbol{\sigma}_z,\\
\H_{\rm Z}&=&\frac{|e|\hbar}{2m}B\boldsymbol{\sigma}_z,\\
\H_{\rm SO}&=&-\gamma\omega a\Bigl[
(-i\hbar\partial_z)\boldsymbol{\sigma}_\varphi-
\bigl(
-i\hbar a^{-1}\partial_\varphi+\frac 12|e|Ba\bigr)
\boldsymbol{\sigma}_z\Bigr]
\end{eqnarray}
with the dimensionless magnetic field $\gamma=\kappa B$ and
\begin{equation}
\boldsymbol{\sigma}_\varphi=\begin{pmatrix}
0&-ie^{-i\varphi}\\ie^{i\varphi} & 0
\end{pmatrix}.
\end{equation}

Let us now write explicitly the effect of these terms in the cylindrical geometry,  acting on a two-component spinor of the form
\begin{equation}
\Psi(\varphi,z)=\begin{pmatrix}\alpha e^{-i\varphi/2}\\ \beta e^{i\varphi/2}
\end{pmatrix}e^{i\ell\varphi} e^{ikz}, \label{eq-spinor}
\end{equation}
with $\alpha$ and $\beta$ constants and
with $\ell\in \mathbb Z$ if we require the fermionic property  under $2\pi$ rotation, $\Psi(\varphi+2\pi,z)=-\Psi(\varphi,z)$~\cite{doi:10.1119/1.4955153}.
\begin{eqnarray}
\H_{\rm K}\Psi&=&\frac 1{2m}
\begin{pmatrix}
\left[
\left(
\frac\hbar a(\ell-1/2)+\frac 12|e|Ba
\right)^2+\hbar^2k^2-\frac 14\frac{\hbar^2}{a^2}\right]\alpha e^{-i\varphi/2}\\
\left[
\left(
\frac\hbar a(\ell+1/2)+\frac 12|e|Ba
\right)^2+\hbar^2k^2-\frac 14\frac{\hbar^2}{a^2}\right] \beta e^{i\varphi/2}\\
\end{pmatrix}
e^{i\ell\varphi} e^{ikz},\\
\H_{\rm I}\Psi&=&-\omega a
\begin{pmatrix}
\left[
\frac\hbar a(\ell-1/2)+\frac 12|e|Ba+\frac 12\frac\hbar a
\right]\alpha e^{-i\varphi/2}\\
\left[\frac\hbar a(\ell+1/2)+\frac 12|e|Ba+\frac 12\frac\hbar a
\right] \beta e^{i\varphi/2}\\
\end{pmatrix} e^{i\ell\varphi} e^{ikz}
,\\
\H_{\rm Z}\Psi&=&
\frac 12\hbar\omega_c\begin{pmatrix}\alpha e^{-i\varphi/2}\\ -\beta e^{i\varphi/2}
\end{pmatrix}e^{i\ell\varphi} e^{ikz},\\
\H_{\rm SO}\Psi&=&
-\gamma\omega a\begin{pmatrix}
 \Bigr[- \left(
\frac\hbar a(\ell-1/2)+\frac 12|e|Ba
\right) \alpha -i\hbar k \beta \Bigl] e^{-i\varphi/2}\\
 \Bigr[i\hbar k \alpha+\left(
\frac\hbar a(\ell+1/2)+\frac 12|e|Ba
\right) \beta \Bigl] e^{i\varphi/2}
\end{pmatrix}e^{i\ell\varphi} e^{ikz}.
\end{eqnarray}

It is worth noticing that a cancellation of the effect of the Zeeman term by the spin-rotation coupling requires that
$\omega=|e|B/m=\omega_c$. The difference by a factor of 2 between the rotation frequency needed to counterbalance orbital and Zeeman effects is due to the Land\'e factor of the electron, here approximated to $g_e\simeq 2$. 
In which concerns the last term in (\ref{eq-Htot}), we observe on the other hand that there is no rotation to compensate for the Pauli spin-orbit term.
In the general case, the spin-orbit interaction mixes the spinor components.
We introduce the following notations for convenience:
\begin{eqnarray}
\hbar k^\sigma_\varphi&=&\frac \hbar a(\ell+\sigma/2)+\frac 12|e|Ba=\frac \hbar a\left(\ell+\sigma/2
+\Phi/\Phi_0\right),\\
\hbar\Omega^\pm&=&\frac{\hbar^2 k^2}{2m}-\frac{\hbar^2}{8ma^2}+
\frac{\hbar^2 (k^\sigma_\varphi)^2}{2m}
-\hbar\omega k^\sigma_\varphi a\pm\frac 12\hbar(\omega-\omega_c)\mp\gamma\hbar\omega k^\sigma_\varphi a,
\end{eqnarray}
with $\Phi_0=2\pi\hbar/|e|\simeq 3.93\ 10^{-15}$USI the flux quantum, and 
in terms of which the eigenvalue equation now reads as
\begin{equation}
\H \begin{pmatrix}\alpha e^{-i\varphi/2}\\ \beta e^{i\varphi/2}
\end{pmatrix}e^{i\ell\varphi} e^{ikz}=
\begin{pmatrix}
\hbar\Omega^- & i\gamma\hbar\omega ka e^{-i\varphi}\\
- i\gamma\hbar\omega ka e^{i\varphi} & \hbar\Omega^+
\end{pmatrix}\begin{pmatrix}\alpha e^{-i\varphi/2}\\ \beta e^{i\varphi/2}
\end{pmatrix}e^{i\ell\varphi} e^{ikz}=E\begin{pmatrix}\alpha e^{-i\varphi/2}\\ \beta e^{i\varphi/2}
\end{pmatrix}e^{i\ell\varphi} e^{ikz}.
\end{equation}

The eigenenergies follow from
\begin{equation}
\left|\begin{matrix}
\hbar\Omega^- -E & i\gamma\hbar\omega ka e^{-i\varphi}\\
- i\gamma\hbar\omega ka e^{i\varphi} & \hbar\Omega^+ -E
\end{matrix}\right|=0
\end{equation}
i.e.
\begin{equation}
E_{\ell k\sigma}=\frac 12\hbar(\Omega^++\Omega^-)+\frac 12\sigma\hbar\sqrt{
(\Omega^+-\Omega^-)^2+4\gamma^2\omega^2k^2a^2 \label{eq-Eellksigma}
}.
\end{equation}

\begin{figure}
	\centering
		%\vspace{-18mm}
        \hspace{-7mm}\includegraphics[scale=1.0]{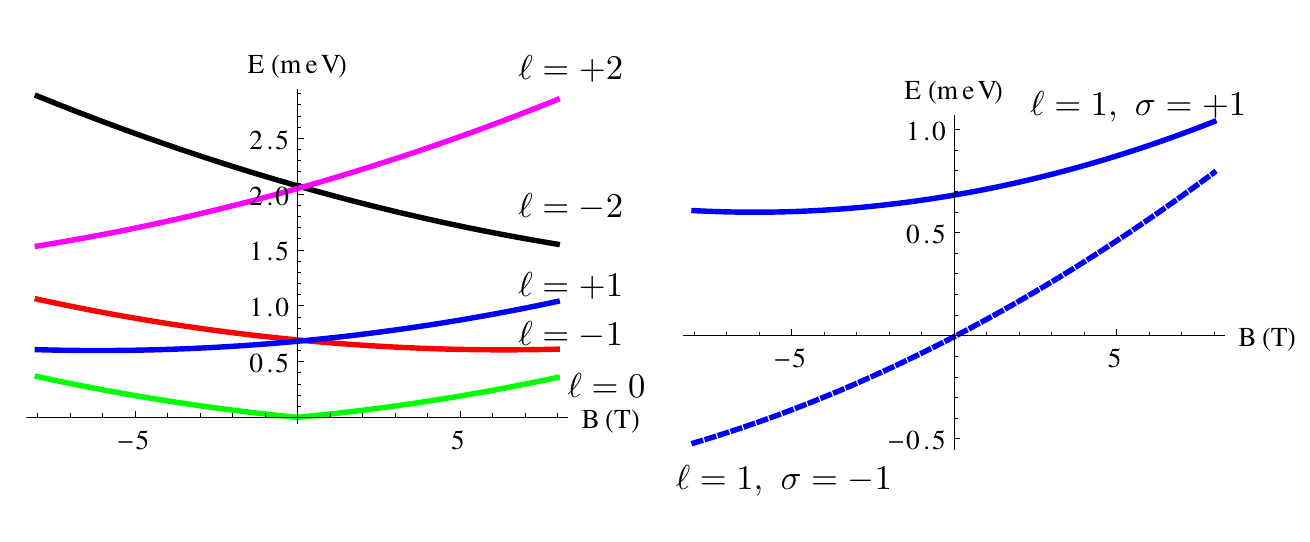}
		%\vspace{-18mm}
		\hspace{-23mm}\includegraphics[scale=1.3]{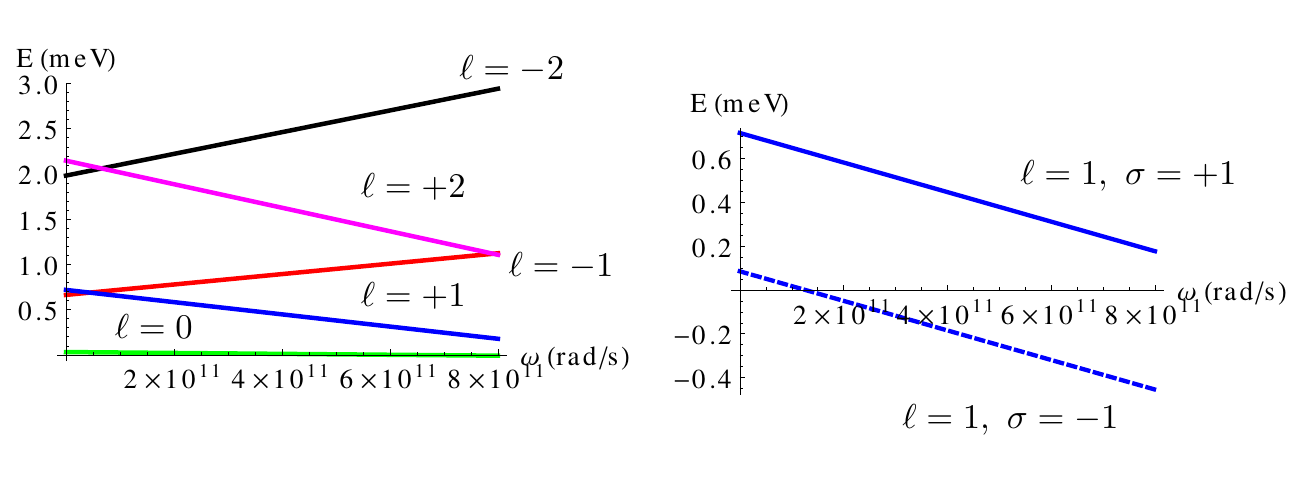}
		%\vspace{-18mm}
        \caption{Energy (in meV), top: as function of the magnetic field, for the axial magnetic field case, when $\omega=10^{10} $rad.s$^{-1}$, bottom: as function of the angular velocity, when $B=1 $T. The values of $\ell$ and $\sigma$ are indicated as plot legends. The radius of the tube is fixed to $a=50$~nm. At that value of the  radius, the ratio $\Phi/\Phi_0$ varies typically from $0$ to $10$ on the scale of the plot. The left plots show the first levels with $\sigma=+1$: $\ell=-2,-1,0,1,2$, $k=0$ (the largest energy scale being $\hbar^2k^2/2m$, we set it to zero to enhance the role of the other parameters), the right column shows for the level $\ell=1$, for the two values of $\sigma=-1,+1$. 
        }\label{Fig1}
\end{figure}

\begin{figure}
\centering\includegraphics[width=\linewidth]{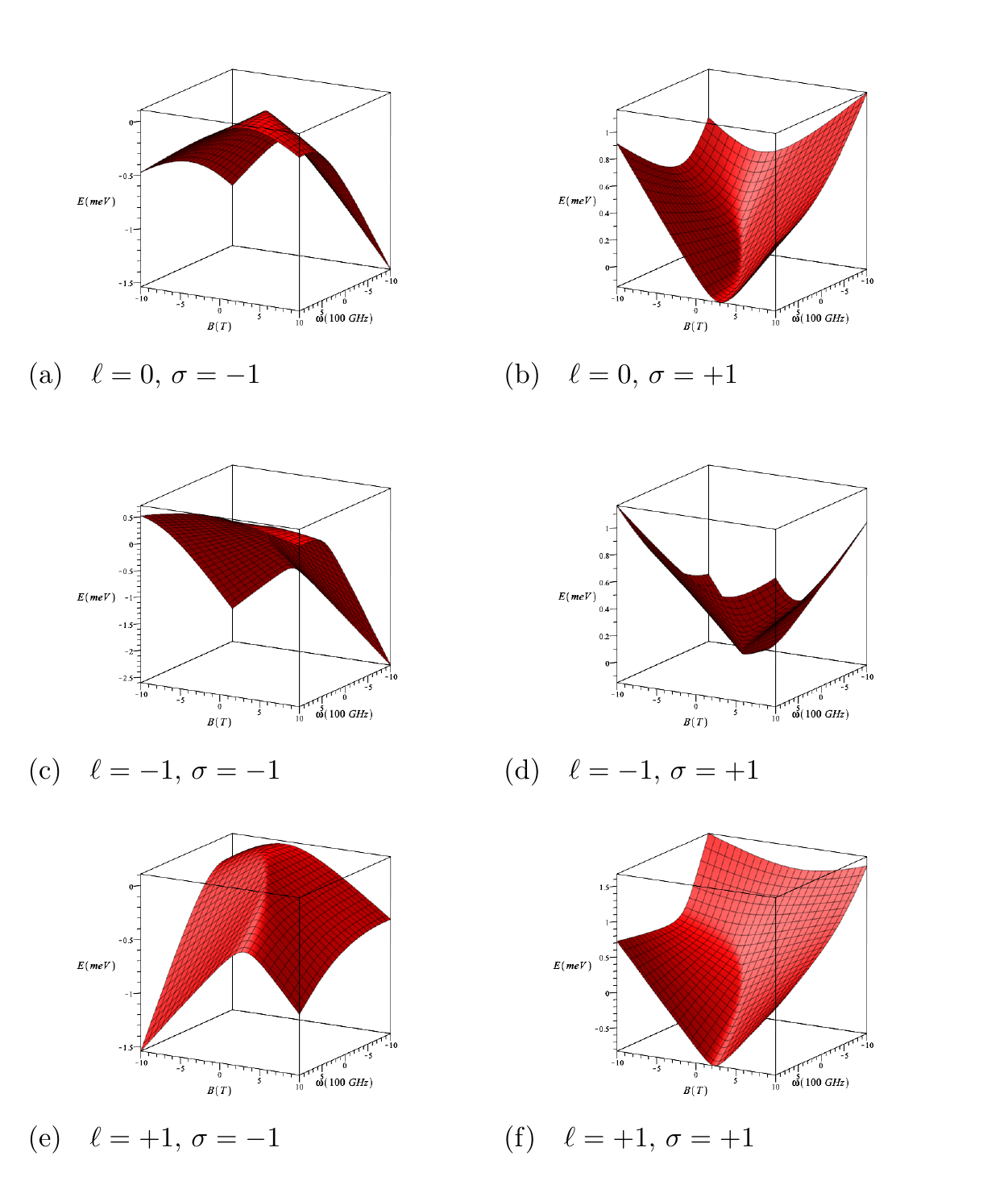}
\caption{Energy (in {\it meV}) of a few $\ket{\ell k \sigma}$ states, in the axial magnetic field case, for $k=1/a$, $a$= 50 nm, as a function of magnetic field  and rotation speed.}
\label{E3d}
\end{figure}

At $\omega=0$, the function $E_{\ell k\sigma}$ with fixed $\sigma$ is periodic in $\Phi/\Phi_0$ (given by a set of parabolas (see FiG.~\ref{Fig1}), top, which satisfy $E_{\ell+n}(\Phi/\Phi_0-n)=E_\ell(\Phi/\Phi_0)$). The spin-orbit interaction lifts the energies degeneracy and the value of the rotation parameter $\omega$ breaks the perfect periodicity in  $\Phi/\Phi_0$.
{The spin-orbit coupling also introduces a combined effect of both rotation and the magnetic field in the term $4\gamma^2 \omega^2k^2a^2$, since $\gamma \omega=\kappa B \omega$}. In Fig. \ref{E3d} it is shown the energy landscape  when $B$ and $\omega$ are varied for a few eigenstates. Notice the correspondence with Fig. \ref{Fig1}.

\subsection{Eigenspinors}

In order to facilitate the search of the eigenspinors, it is worth writing the Hamiltonian under the form
\begin{align}
\H&=\frac 12\hbar(\Omega^-+\Omega^+)\One+\frac 12\hbar(\Omega^--\Omega^+)\boldsymbol{\sigma}_z-\gamma\hbar\omega ka\boldsymbol{\sigma}_\varphi \label{eqH} \\ 
&=\frac 12\hbar(\Omega^-+\Omega^+)\One+\frac 12\hbar(\Omega^--\Omega^+)
\left[
\boldsymbol{\sigma}_z-\frac{2\gamma\omega ka}{\Omega^--\Omega^+}\boldsymbol{\sigma}_\varphi
\right] \nonumber
\end{align}
where the last bracket can also be denoted as
\begin{equation}
{ \boldsymbol{\tilde\sigma}}_\varphi=\boldsymbol{\sigma}_z-\tan\theta\boldsymbol{\sigma}_\varphi,\qquad\tan\theta=\frac{2\gamma\omega ka}{\Omega^--\Omega^+}. \label{tan}
\end{equation}

The normalized eigenstates  $\Psi_{\ell k \sigma}$  of ${ \boldsymbol{\tilde\sigma}}_\varphi$, hence of $\H$, are
\begin{eqnarray}\Psi_{\ell k +}&=
\begin{pmatrix}
\cos\frac{\theta}{2}e^{-i\varphi/2}\\ -i\sin\frac{\theta}{2}e^{i\varphi/2}
\end{pmatrix}e^{i(\ell\varphi+kz)},\quad
\Psi_{\ell k -}&=
\begin{pmatrix}
i\sin\frac{\theta}{2}e^{-i\varphi/2}\\ -\cos\frac{\theta}{2}e^{i\varphi/2}
\end{pmatrix}e^{i(\ell\varphi+kz)}.\label{Eq-eigenstates}
\end{eqnarray}

A word of caution is needed here. Although the transformation given by Eq. (\ref{tan}) provides an elegant way of presenting the eigenstates, it is singular at $\Omega^-  =\Omega^+$ and therefore not valid when this happens. As $\Omega^-$ approaches $\Omega^+$ from below and goes above it, $\tan\theta$ jumps from $-\infty$ to $+\infty$ which, obviously, is not physical since there is no such jump in Eq. (\ref{eqH}). Further, $\boldsymbol{\tilde\sigma}_\varphi$ is meaningless in this case. Of course this is just an artifact of the notation which   was carefully taken into consideration when plotting the charge and spin currents, which explicitly depend on $\sin\theta$ and $\cos\theta$. 

\subsection{Orders of magnitude}

Carbon nanotubes are good candidates to analyse quantitatively the effect of simultaneous presence of rotation and magnetic field. They have various electronic structures (metallic or semiconductor along the axis) depending on their chirality.  Typical order of magnitude for a  carbon nanotube diameter is $a\simeq 1-500$~nm. In our study we fix $a=50$~nm. Typical laboratory magnetic fields are of the order 1~T which gives a corresponding cyclotron frequency of order 100~GHz, which is compatible with the nanomotors rotation frequency mentioned at the introduction.

%\begin{figure}
%	\centering
%		%\vspace{-18mm}
%        \hspace{-7mm}\includegraphics[scale=0.4]{E-.png}
		%\vspace{-18mm}
%		\hspace{-8mm}\includegraphics[scale=0.4]{E+.png}
			%\vspace{-18mm}
 %       \caption{Energy ( in meV), for the axial magnetic field case, for $k=1/a$, $a$= 50 nm, $\ell=0$, as a function of magnetic field  and rotation speed.  Left, for $\sigma=-1$; right, for $\sigma=+1$.
 %       }\label{E3d}
%\end{figure}
%\begin{figure}
%	\centering
		%\vspace{-18mm}
 %       \hspace{-7mm}\includegraphics[scale=0.4]{E-1-.png}
		%\vspace{-18mm}
%		\hspace{-8mm}\includegraphics[scale=0.4]{E+1-.png}
		%\vspace{-18mm}
 %       \caption{Energy ( in meV), for the axial magnetic field case, for $k=1/a$, $a$= 50 nm, $\ell=-1$, as a function of magnetic field  and rotation speed.  Left, for $\sigma=-1$; right, for $\sigma=+1$.
  %      }\label{E3d2}
%\end{figure}
%\begin{figure}
%	\centering
		%\vspace{-18mm}
 %       \hspace{-7mm}\includegraphics[scale=0.4]{E-1+.png}
		%\vspace{-18mm}
%		\hspace{-8mm}\includegraphics[scale=0.4]{E+1+.png}
		%\vspace{-18mm}
 %       \caption{Energy ( in meV), for the axial magnetic field case, for $k=1/a$, $a$= 50 nm, $\ell=1$, as a function of magnetic field  and rotation speed.  Left, for $\sigma=-1$; right, for $\sigma=+1$.
  %      }\label{E3d2}
%\end{figure}

\section{Charge and spin current densities}

%\subsection{Axial magnetic field}

\subsection{Charge currents}
We will now focus on the charge currents in a pure quantum state,  and in the next section, on spin current at $T=0$.
For a given energy channel, the charge current density at $T=0$ can be calculated  using the  definition
\begin{equation}
\vec{J}_{\ell k \sigma}=-|e|\Psi^\dagger_{\ell k \sigma}\vec{\bf v}\Psi_{\ell k \sigma}\quad\hbox{with}\quad\vec{\mathbf v}=\frac{i}{\hbar}[\H,\vec{r}\ \!].\label{eq-Jcharge}
\end{equation}
Following Ref. \cite{0143-0807-34-1-161}, we have defined the charge current in  such way that the dimension of $\vec{J}$  is that of  charge times  velocity. Since the motion is constrained to the nanotube, only two spatial components of $\vec{\mathbf v}$ are needed,
${\bf v}_\varphi=\frac{ia}{\hbar}[\H,\varphi]$
and ${\bf v}_z=\frac{i}{\hbar}[\H,z]$. The calculation gives
\begin{equation}
{\bf v}_\varphi=\frac{1}{m}\Bigl(-i\hbar a^{-1}\partial_\varphi+\frac{|e|Ba}{2}\Bigr)\One -\omega a\One +\gamma\omega a\boldsymbol{\sigma}_z\label{eq-vvarphi}
\end{equation}
and
\begin{equation}
{\bf v}_z=\frac{1}{m}(-i\hbar \partial_z)\One -\gamma\omega a\boldsymbol{\sigma}_\varphi.\label{eq-vz}
\end{equation}
The charge current density in the azimuthal direction follows,
\begin{equation}
J_{\varphi,\ell k \sigma}=-|e|\Bigl[
\frac{\hbar\ell}{ma}+\frac{|e|Ba}{2m}-\omega a+\sigma\bigl(\gamma\omega a-\frac\hbar{2ma}\bigr)\cos\theta
\Bigr].\label{eq-Jvarphi}
\end{equation}
The first term $\propto\hbar\ell$ is the paramagnetic current density while the second term, linear in $B$, is the diamagnetic contribution. The next term is its rotation counterpart. The last term, depending on $\gamma$ is due to the spin-orbit interaction. As discussed in the beginning of the paper, when $\omega=\frac 12\omega_c$, the orbital contributions of the magnetic field and of rotation cancel each other.
%\begin{figure}[!h!]
%\centering
%\includegraphics[width=0.5\textwidth]{plot3D_Jphi_axial.pdf}
%\caption{Plot of the $J_{\varphi} \times 10^{15}$ as a function of the magnetic field and rotation, for the case of axial field and $\sigma=+1$.}
%\label{Jphi_axial_3D}
%\end{figure}
In the nanotube axis direction, the charge current density is
\begin{equation}
J_{z,\ell k \sigma}=-|e|\Bigl[
\frac{\hbar k}{m}+\sigma\gamma\omega a\sin\theta
\Bigr].
\label{eq-Jz_axial}
\end{equation}

The term $\propto\hbar k$ is also a paramagnetic contribution  due to the ``initial conditions''  in the selection of the eigenstate while the second term, with $\gamma$ dependence, results from SO interaction. Note that $\omega$, $B$ and the SO interaction appear on  both components.

An interesting issue here concerns the definition of the charge current from the Lagrangian approach. The Lagrangian can be written, in a state $\Psi$ as~\cite{0143-0807-31-5-026}
\be
L=\<\Psi|i\hbar\partial_t\One-\H|\Psi\>.
\ee
Here we only deal with stationary states, which simplifies the expression, and, using the definition of the current in terms of $L$, one has~\cite{0143-0807-34-1-161}
\be\vec j=\frac{\partial L}{\partial\vec A}=-\frac{\partial}{\partial\vec A}\<\Psi|\H|\Psi\>,\ee
but there are caveats here: first the energy has to be expressed in terms of the vector potential and not  as a function of the magnetic field (for example the Zeeman term does not couple spin to $\vec A$, but to $\vec B$ directly), second this approach does not allow to define the current in the $z$ direction (since $\vec A$ has no component along $z$). The calculation in the eigenstates (\ref{Eq-eigenstates}) leads to
\be
-\frac{\partial}{\partial A_\varphi}E_{\ell k \sigma}=-|e|\Bigl(
\frac{\hbar\ell}{ma}+\frac{|e|A_\varphi}{m}-\omega a+\sigma\bigl(\gamma\omega a-\frac\hbar{2ma}\bigr)\cos\theta
\Bigr),
\ee
which identifies to $J_{\varphi,\ell k\sigma}$ according to (\ref{eq-Jvarphi}). As noticed, we do not get  (\ref{eq-Jz_axial}) directly.
On the other hand, it is easy to choose another gauge for the vector potential which leads to the same magnetic field, $\vec A=\frac 12Ba\hat\varphi+A_z\hat z$ with constant $A_z$, which modifies the kinetic term according to
\begin{equation}
\frac 1{2m}(-i\hbar\partial_z)^2\longrightarrow \frac 1{2m}(-i\hbar\partial_z+|e|A_z)^2
\end{equation}
and the SO term according to
\begin{equation}
\H_{\rm SO}
\longrightarrow 
\H_{\rm SO}+\omega aB|e|A_z\boldsymbol{\sigma}_\varphi.\end{equation}
This would lead to reparametrization of the $\Omega$'s, $\theta$, etc. but would allow the calculation of the $z-$current through the formula $J_{z,\ell k \sigma}=-\frac{\partial E'_{\ell k\sigma}}{\partial A_z}$.

Another feature of this expression of the current density is that there is no contribution of the Zeeman current, although the magnetic field is involved. Such a term, associated to the spin polarization~\cite{doi:10.1119/1.4868094}
\be
\frac{-|e|}m\vec\nabla\times(\Psi^\dagger\vec{\mathbf s}\Psi)
\ee
indeed vanishes here due to the uniform character of the spin density.

\subsection{Spin currents}
The spin current density in an eigenstate obeys a  definition similar to (\ref{eq-Jcharge}) (see e.g. Ref.~\cite{0143-0807-31-5-026,0143-0807-34-1-161}),
\begin{equation}
\vec{S}_{\ell k \sigma}^a=\frac 12\Psi^\dagger_{\ell k \sigma}\{\vec{\mathbf{v}},\mathbf s_a\}\Psi_{\ell k \sigma}\quad\hbox{with}\quad s_a=\frac 12\hbar\sigma_a 
\end{equation}
with the velocities given in (\ref{eq-vvarphi}) and (\ref{eq-vz})
and where the anticommutator is required for symmetrization. We use   $S$ to denote the spin current density and the tensorial character is encoded in the upperscript $a$ which refers to the spin polarization considered.
In the azimuthal ($\varphi$) direction, we have, for the two spin labels
\begin{equation}
{S}^z_{\varphi,\ell k \sigma}=\frac{\hbar}{2} \Biggl[\sigma
\Biggl(\frac{\hbar\ell}{ma}+\frac{|e| Ba}{2m}-\omega a\Biggr)\cos\theta
+\gamma\omega a-\frac\hbar{2ma}
\Biggr]. \label{spincphi}
\end{equation}
In the $z$ direction, we have
\begin{equation}
{S}^z_{z,\ell k \sigma}=\frac{\hbar}{2} \Biggl[\sigma\frac{\hbar k}{m}
\cos\theta \Biggr]. \label{spincz}
\end{equation} 
Again, we note that the   spin current depends on both the magnetic field intensity and the rotation velocity. But, differently from the charge current, the SO term  contributes only to the $\varphi$ component. Moreover, a comparison between (\ref{eq-Jvarphi}) and (\ref{spincphi}) and between (\ref{eq-Jz_axial}) and (\ref{spincz}) shows that, for a given eigenstate, it is possible to tune either the magnetic field and/or the rotation velocity in order to cancel the charge currents while keeping non-vanishing spin currents. For instance, for the $z$-component of the charge current this happens provided that 
\begin{equation}
    \gamma  \omega\sin\theta=-\sigma\frac{\hbar k}{ma} . \label{tuning-para}
\end{equation}
On the other hand, the cancellation of the $z$-component of spin current happens at combinations of $\omega$ and $B$ such that
\begin{equation}
    \cos\theta =0 .
\end{equation}
In this case,  the $z$-component of the spin current  vanishes, leaving a charge only current that depends both on the magnetic field and rotation speed and, amazingly, on the spin polarization state, as can be seen in Eq. (\ref{eq-Jz_axial}). This is due to the SO term that couples spin polarization, magnetic field and rotation. Inspection of Eq. (\ref{tan}) shows that this case corresponds to having $\Omega^+ = \Omega^-$, which gives a simpler relation between $B$ and $\omega$, that is $2\kappa(\ell +\frac{\sigma}{2})B\omega +|e| a^2 B^2 \omega + \frac{|e|}{m}B -\omega=0$, besides simplification of the eigenstate energy, Eq. (\ref{eq-Eellksigma}).

\begin{figure}
\centering\includegraphics[width=\linewidth]{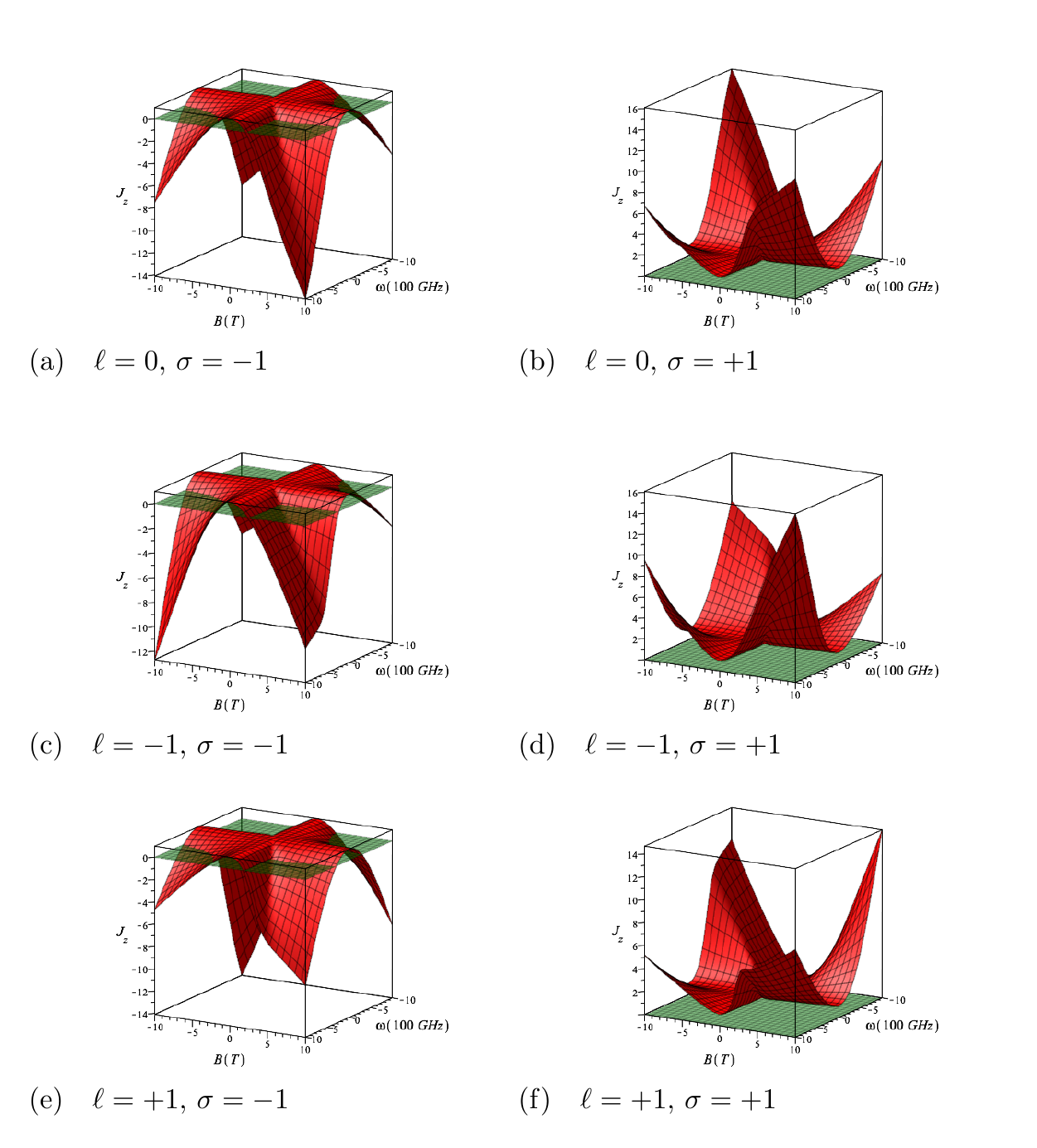}
\caption{The axial component of the charge current (in units of $\frac{e\hbar}{ma}$) of a few $\ket{\ell k\sigma}$ states, in the axial magnetic field case, for $k=1/a$, $a$= 50 nm, as a function of magnetic field  and rotation speed.}
\label{chargecurr}
\end{figure}

\begin{figure}
\centering\includegraphics[width=\linewidth]{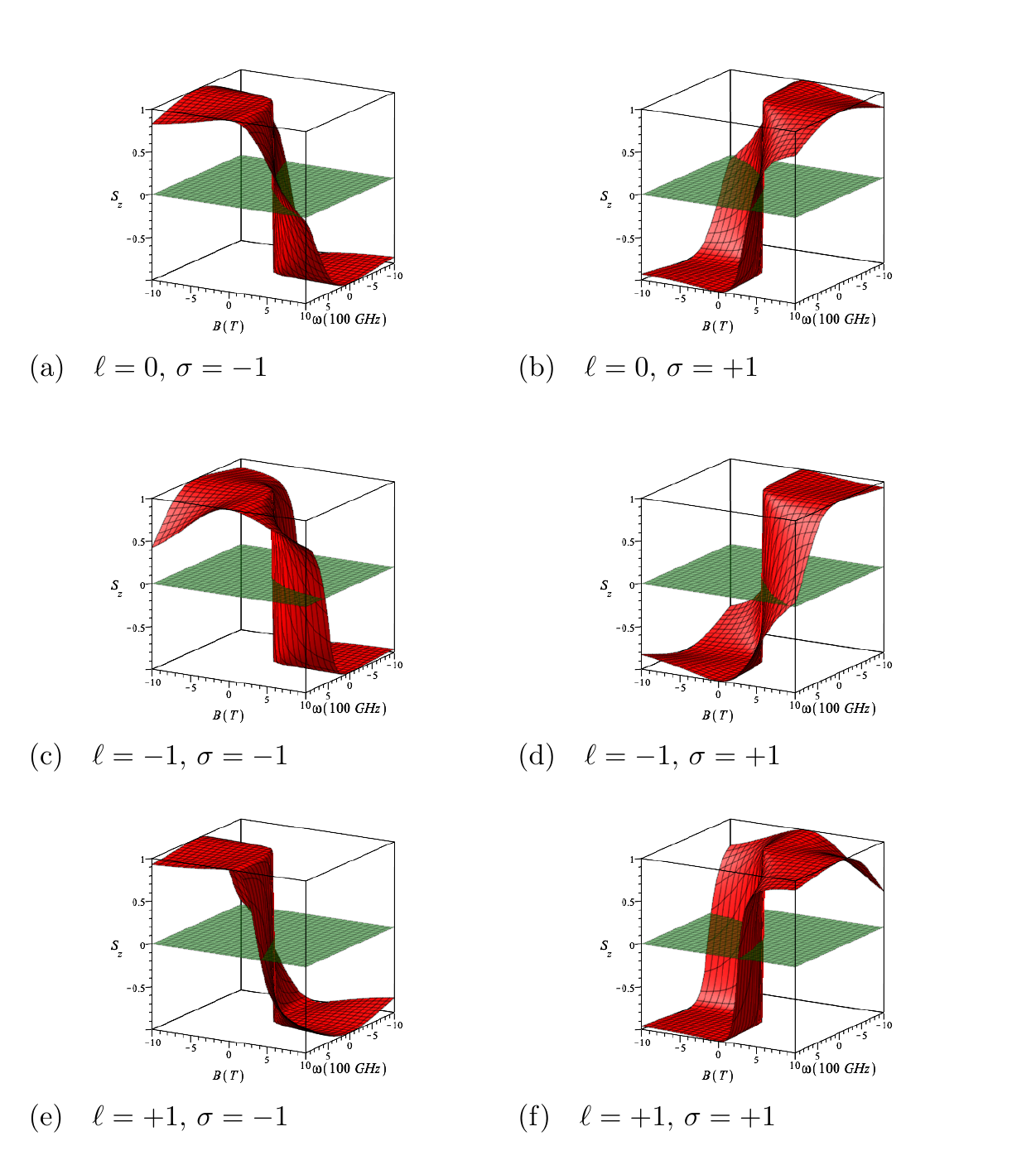}
\caption{The axial component of the spin current (in units of $\frac{\hbar^2}{ma}$) of a few $\ket{\ell k \sigma}$ states, in the axial magnetic field case, for $k=1/a$, $a$= 50 nm, as a function of magnetic field  and rotation speed.}
\label{spincurr}
\end{figure}

\begin{figure}
\centering\includegraphics[width=\linewidth]{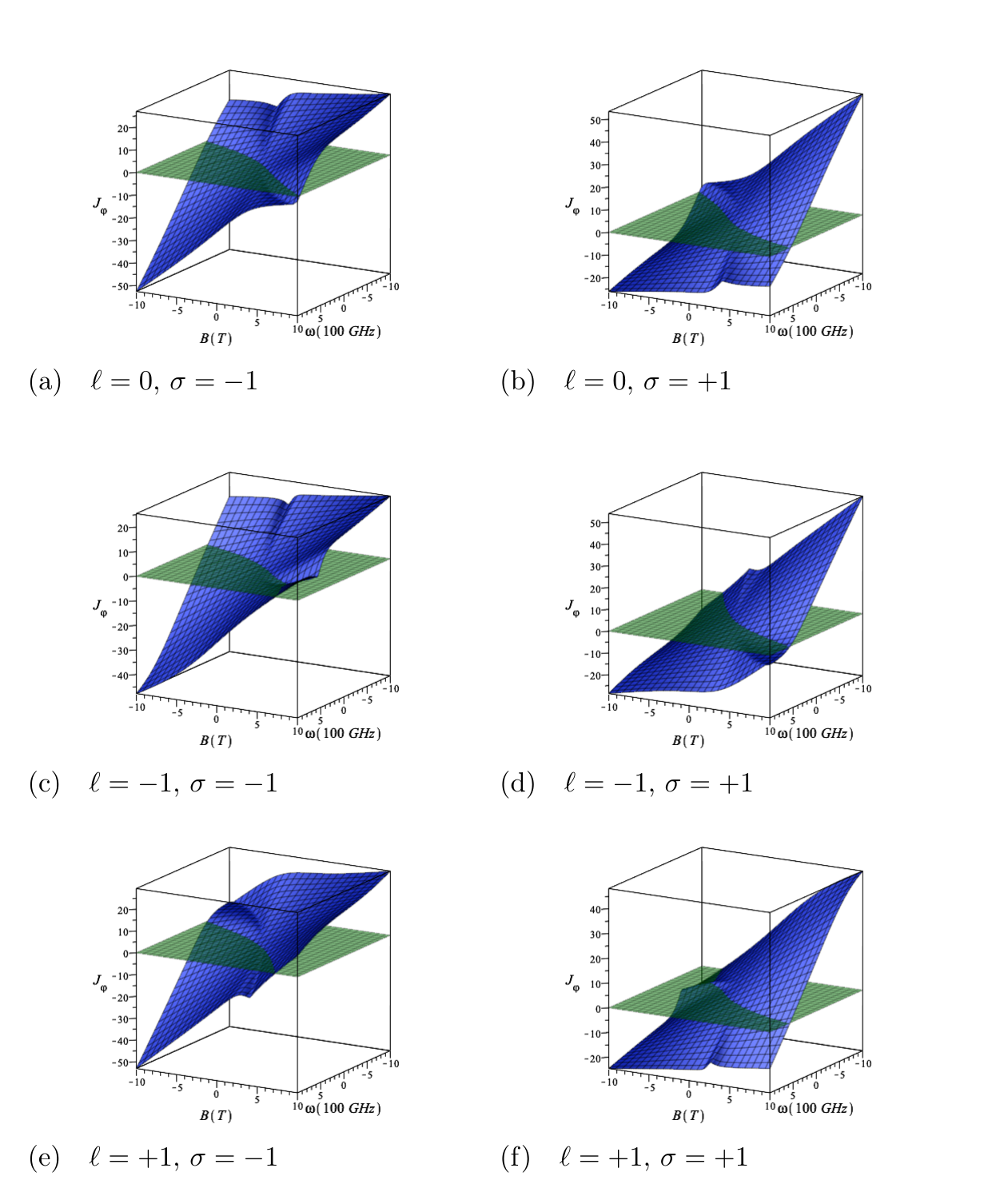}
\caption{The azimuthal component of the charge current (in units of $\frac{e\hbar}{ma}$) of a few $\ket{\ell k\sigma}$ states, in the axial magnetic field case, for $k=1/a$, $a$= 50 nm, as a function of magnetic field  and rotation speed.}
\label{chargecurrphi}
\end{figure}

\begin{figure}
\centering\includegraphics[width=\linewidth]{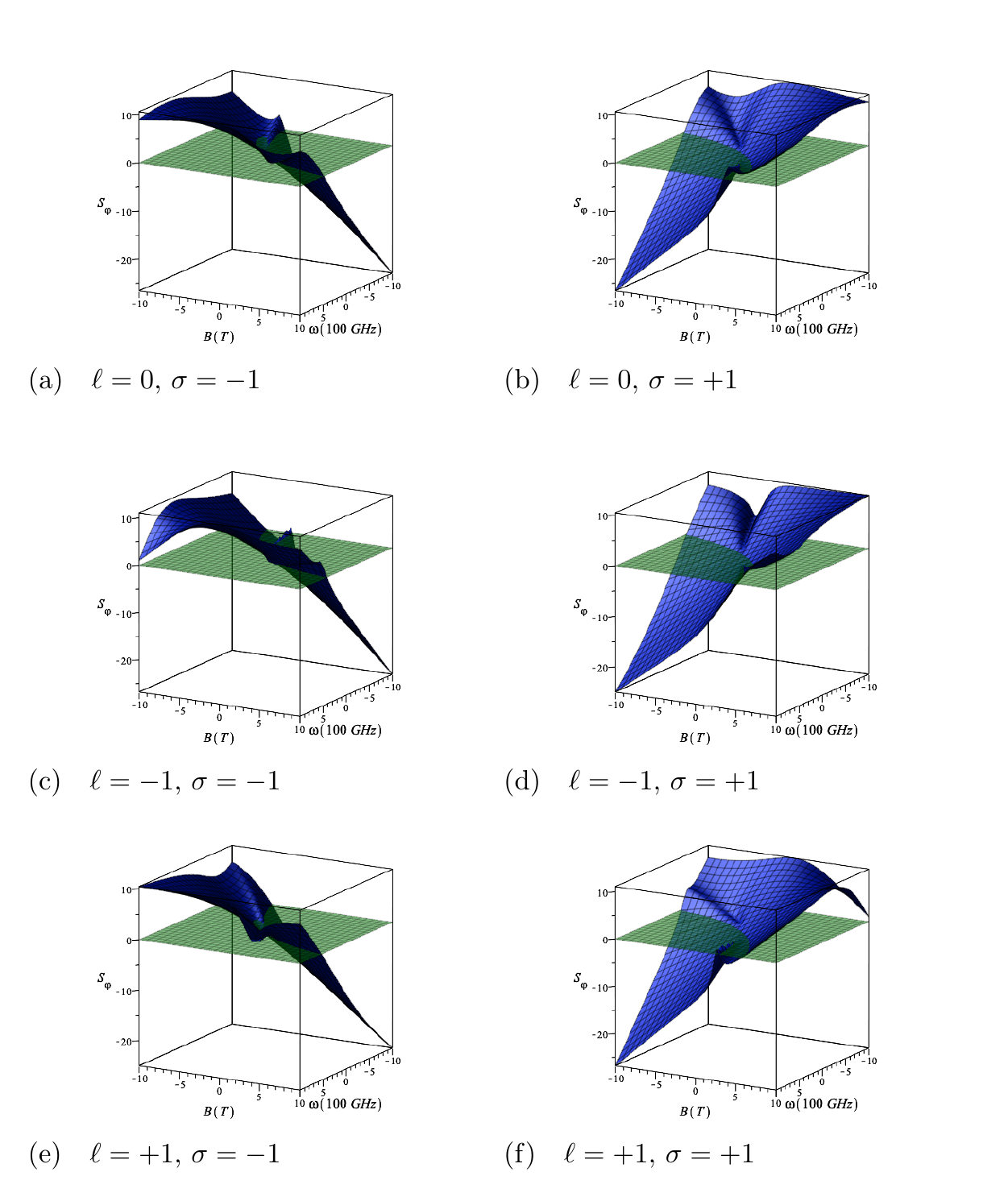}
\caption{The azimuthal component of the spin current (in units of $\frac{\hbar^2}{ma}$) of a few $\ket{\ell k\sigma}$ states, in the axial magnetic field case, for $k=1/a$, $a$= 50 nm, as a function of magnetic field  and rotation speed.}
\label{spincurrphi}
\end{figure}

The control over which component of either current is tuned to zero is evident in the plots of the currents shown in Figs. \ref{chargecurr} and \ref{spincurr}, for the axial components, and  Figs. \ref{chargecurrphi} and \ref{spincurrphi}, for the azimuthal components. Furthermore, those figures also show that the currents might have their direction inverted by choice of the appropriate sector of parameter space ($B$, $\omega$). This provides an effective way of controlling the balance between charge and spin currents and their respective directions.

\section{Conclusion}

In the previous sections we studied electronic and spintronic effects on  ballistic electrons in a rotating nanotube under an applied magnetic field. For the case of an axial magnetic field, we found the eigenenergies and showed that spin and charge currents are generated by ballistic charge injection in the nanotube.  The combined inertial and electromagnetic effects allows for  control of the balance between the charge and spin currents. It is shown that, by playing with the applied field, rotation speed and injection momentum, one can tune the charge current to vanishing values, leaving a nonzero spin current.

At sufficiently low temperatures, electrons in most metals and alloys (provided they have no magnetic order) exhibit a Fermi liquid behaviour with renormalized parameters such as the effective mass \cite{Phillips}. Hence, for a conducting nanotube, the set of weakly-interacting electrons maintained in one of the eigenstates can in principle lead to tunable pure spin currents along the tube axis. Moreover, another possible application is to use the device either as an accelerometer or as a magnetometer: indeed, once the device is tuned such that charge currents are canceled, any change in the magnetic field $\Delta B$ or in the rotation velocity $\Delta \omega$ will break condition (\ref{tuning-para}). Thus, a charge current is generated, the intensity of which can be used to retrieve $\Delta B$ or $\Delta \omega$.

The great difficulty of injecting a ballistic current into a rotating nanotube via physically contacted electrodes can be overcome by  photocurrent injection at optical frequencies \cite{newson2011coherent}. Although our results were obtained for DC currents they can be extended to the AC domain, which is the aim of future work.

\textbf{Acknowledgments}: This work was partially supported by the Brazilian agencies CAPES, CNPq, INCT-nanocarbono and FACEPE and by the German agency Alexander von Humboldt Foundation. MMC acknowledges CAPES Grant 88887.358036/2019-00.

%merlin.mbs apsrev4-1.bst 2010-07-25 4.21a (PWD, AO, DPC) hacked
%Control: key (0)
%Control: author (8) initials jnrlst
%Control: editor formatted (1) identically to author
%Control: production of article title (-1) disabled
%Control: page (0) single
%Control: year (1) truncated
%Control: production of eprint (0) enabled
%

%\bibliography{bib} 

\begin{thebibliography}{33}%
\makeatletter
\providecommand \@ifxundefined [1]{%
 \@ifx{#1\undefined}
}%
\providecommand \@ifnum [1]{%
 \ifnum #1\expandafter \@firstoftwo
 \else \expandafter \@secondoftwo
 \fi
}%
\providecommand \@ifx [1]{%
 \ifx #1\expandafter \@firstoftwo
 \else \expandafter \@secondoftwo
 \fi
}%
\providecommand \natexlab [1]{#1}%
\providecommand \enquote  [1]{``#1''}%
\providecommand \bibnamefont  [1]{#1}%
\providecommand \bibfnamefont [1]{#1}%
\providecommand \citenamefont [1]{#1}%
\providecommand \href@noop [0]{\@secondoftwo}%
\providecommand \href [0]{\begingroup \@sanitize@url \@href}%
\providecommand \@href[1]{\@@startlink{#1}\@@href}%
\providecommand \@@href[1]{\endgroup#1\@@endlink}%
\providecommand \@sanitize@url [0]{\catcode `\\12\catcode `\$12\catcode
  `\&12\catcode `\#12\catcode `\^12\catcode `\_12\catcode `\%12\relax}%
\providecommand \@@startlink[1]{}%
\providecommand \@@endlink[0]{}%
\providecommand \url  [0]{\begingroup\@sanitize@url \@url }%
\providecommand \@url [1]{\endgroup\@href {#1}{\urlprefix }}%
\providecommand \urlprefix  [0]{URL }%
\providecommand \Eprint [0]{\href }%
\providecommand \doibase [0]{http://dx.doi.org/}%
\providecommand \selectlanguage [0]{\@gobble}%
\providecommand \bibinfo  [0]{\@secondoftwo}%
\providecommand \bibfield  [0]{\@secondoftwo}%
\providecommand \translation [1]{[#1]}%
\providecommand \BibitemOpen [0]{}%
\providecommand \bibitemStop [0]{}%
\providecommand \bibitemNoStop [0]{.\EOS\space}%
\providecommand \EOS [0]{\spacefactor3000\relax}%
\providecommand \BibitemShut  [1]{\csname bibitem#1\endcsname}%
\let\auto@bib@innerbib\@empty
%</preamble>
\bibitem [{\citenamefont {Cao}\ \emph {et~al.}(2019)\citenamefont {Cao},
  \citenamefont {Cong}, \citenamefont {Cao}, \citenamefont {Wu}, \citenamefont
  {Liu}, \citenamefont {Amer},\ and\ \citenamefont {Zhou}}]{cao2019review}%
  \BibitemOpen
  \bibfield  {author} {\bibinfo {author} {\bibfnamefont {Y.}~\bibnamefont
  {Cao}}, \bibinfo {author} {\bibfnamefont {S.}~\bibnamefont {Cong}}, \bibinfo
  {author} {\bibfnamefont {X.}~\bibnamefont {Cao}}, \bibinfo {author}
  {\bibfnamefont {F.}~\bibnamefont {Wu}}, \bibinfo {author} {\bibfnamefont
  {Q.}~\bibnamefont {Liu}}, \bibinfo {author} {\bibfnamefont {M.~R.}\
  \bibnamefont {Amer}}, \ and\ \bibinfo {author} {\bibfnamefont
  {C.}~\bibnamefont {Zhou}},\ }in\ \href@noop {} {\emph {\bibinfo {booktitle}
  {Single-Walled Carbon Nanotubes}}}\ (\bibinfo  {publisher} {Springer},\
  \bibinfo {year} {2019})\ pp.\ \bibinfo {pages} {189--224}\BibitemShut
  {NoStop}%
\bibitem [{\citenamefont {Li}\ \emph {et~al.}(2019)\citenamefont {Li},
  \citenamefont {Wang},\ and\ \citenamefont {Shi}}]{LI2019260}%
  \BibitemOpen
  \bibfield  {author} {\bibinfo {author} {\bibfnamefont {Y.}~\bibnamefont
  {Li}}, \bibinfo {author} {\bibfnamefont {A.}~\bibnamefont {Wang}}, \ and\
  \bibinfo {author} {\bibfnamefont {J.}~\bibnamefont {Shi}},\ }\href {\doibase
  https://doi.org/10.1016/j.commatsci.2018.09.046} {\bibfield  {journal}
  {\bibinfo  {journal} {Computational Materials Science}\ }\textbf {\bibinfo
  {volume} {156}},\ \bibinfo {pages} {260 } (\bibinfo {year}
  {2019})}\BibitemShut {NoStop}%
\bibitem [{\citenamefont {Li}\ \emph {et~al.}(2014)\citenamefont {Li},
  \citenamefont {Wang}, \citenamefont {Zhao}, \citenamefont {Gao},
  \citenamefont {Zhao},\ and\ \citenamefont {Zhou}}]{li2014rotation}%
  \BibitemOpen
  \bibfield  {author} {\bibinfo {author} {\bibfnamefont {J.}~\bibnamefont
  {Li}}, \bibinfo {author} {\bibfnamefont {X.}~\bibnamefont {Wang}}, \bibinfo
  {author} {\bibfnamefont {L.}~\bibnamefont {Zhao}}, \bibinfo {author}
  {\bibfnamefont {X.}~\bibnamefont {Gao}}, \bibinfo {author} {\bibfnamefont
  {Y.}~\bibnamefont {Zhao}}, \ and\ \bibinfo {author} {\bibfnamefont
  {R.}~\bibnamefont {Zhou}},\ }\href@noop {} {\bibfield  {journal} {\bibinfo
  {journal} {Scientific Reports}\ }\textbf {\bibinfo {volume} {4}},\ \bibinfo
  {pages} {5846} (\bibinfo {year} {2014})}\BibitemShut {NoStop}%
\bibitem [{\citenamefont {Narendar}\ and\ \citenamefont
  {Gopalakrishnan}(2011)}]{narendar2011nonlocal}%
  \BibitemOpen
  \bibfield  {author} {\bibinfo {author} {\bibfnamefont {S.}~\bibnamefont
  {Narendar}}\ and\ \bibinfo {author} {\bibfnamefont {S.}~\bibnamefont
  {Gopalakrishnan}},\ }\href@noop {} {\bibfield  {journal} {\bibinfo  {journal}
  {Results in Physics}\ }\textbf {\bibinfo {volume} {1}},\ \bibinfo {pages}
  {17} (\bibinfo {year} {2011})}\BibitemShut {NoStop}%
\bibitem [{\citenamefont {Belhadj}\ \emph {et~al.}(2017)\citenamefont
  {Belhadj}, \citenamefont {Boukhalfa},\ and\ \citenamefont
  {Belalia}}]{belhadj2017free}%
  \BibitemOpen
  \bibfield  {author} {\bibinfo {author} {\bibfnamefont {A.}~\bibnamefont
  {Belhadj}}, \bibinfo {author} {\bibfnamefont {A.}~\bibnamefont {Boukhalfa}},
  \ and\ \bibinfo {author} {\bibfnamefont {S.~A.}\ \bibnamefont {Belalia}},\
  }\href@noop {} {\bibfield  {journal} {\bibinfo  {journal} {The European
  Physical Journal Plus}\ }\textbf {\bibinfo {volume} {132}},\ \bibinfo {pages}
  {513} (\bibinfo {year} {2017})}\BibitemShut {NoStop}%
\bibitem [{\citenamefont {Wang}\ \emph {et~al.}(2017)\citenamefont {Wang},
  \citenamefont {Wu},\ and\ \citenamefont {Wang}}]{wang2017design}%
  \BibitemOpen
  \bibfield  {author} {\bibinfo {author} {\bibfnamefont {L.}~\bibnamefont
  {Wang}}, \bibinfo {author} {\bibfnamefont {H.}~\bibnamefont {Wu}}, \ and\
  \bibinfo {author} {\bibfnamefont {F.}~\bibnamefont {Wang}},\ }\href@noop {}
  {\bibfield  {journal} {\bibinfo  {journal} {Scientific reports}\ }\textbf
  {\bibinfo {volume} {7}},\ \bibinfo {pages} {41717} (\bibinfo {year}
  {2017})}\BibitemShut {NoStop}%
\bibitem [{\citenamefont {Tu}\ \emph {et~al.}(2016)\citenamefont {Tu},
  \citenamefont {Yang}, \citenamefont {Wang},\ and\ \citenamefont
  {Li}}]{tu2016rotating}%
  \BibitemOpen
  \bibfield  {author} {\bibinfo {author} {\bibfnamefont {Q.}~\bibnamefont
  {Tu}}, \bibinfo {author} {\bibfnamefont {Q.}~\bibnamefont {Yang}}, \bibinfo
  {author} {\bibfnamefont {H.}~\bibnamefont {Wang}}, \ and\ \bibinfo {author}
  {\bibfnamefont {S.}~\bibnamefont {Li}},\ }\href@noop {} {\bibfield  {journal}
  {\bibinfo  {journal} {Scientific reports}\ }\textbf {\bibinfo {volume} {6}},\
  \bibinfo {pages} {26183} (\bibinfo {year} {2016})}\BibitemShut {NoStop}%
\bibitem [{\citenamefont {{\v{Z}}uti{\'c}}\ \emph {et~al.}(2004)\citenamefont
  {{\v{Z}}uti{\'c}}, \citenamefont {Fabian},\ and\ \citenamefont
  {Sarma}}]{vzutic2004spintronics}%
  \BibitemOpen
  \bibfield  {author} {\bibinfo {author} {\bibfnamefont {I.}~\bibnamefont
  {{\v{Z}}uti{\'c}}}, \bibinfo {author} {\bibfnamefont {J.}~\bibnamefont
  {Fabian}}, \ and\ \bibinfo {author} {\bibfnamefont {S.~D.}\ \bibnamefont
  {Sarma}},\ }\href@noop {} {\bibfield  {journal} {\bibinfo  {journal} {Reviews
  of modern physics}\ }\textbf {\bibinfo {volume} {76}},\ \bibinfo {pages}
  {323} (\bibinfo {year} {2004})}\BibitemShut {NoStop}%
\bibitem [{\citenamefont {Joshi}(2016)}]{joshi2016spintronics}%
  \BibitemOpen
  \bibfield  {author} {\bibinfo {author} {\bibfnamefont {V.~K.}\ \bibnamefont
  {Joshi}},\ }\href@noop {} {\bibfield  {journal} {\bibinfo  {journal}
  {Engineering science and technology, an international journal}\ }\textbf
  {\bibinfo {volume} {19}},\ \bibinfo {pages} {1503} (\bibinfo {year}
  {2016})}\BibitemShut {NoStop}%
\bibitem [{\citenamefont {Guimaraes}\ \emph {et~al.}(2010)\citenamefont
  {Guimaraes}, \citenamefont {Kirwan}, \citenamefont {Costa}, \citenamefont
  {Muniz}, \citenamefont {Mills},\ and\ \citenamefont
  {Ferreira}}]{guimaraes2010carbon}%
  \BibitemOpen
  \bibfield  {author} {\bibinfo {author} {\bibfnamefont {F.}~\bibnamefont
  {Guimaraes}}, \bibinfo {author} {\bibfnamefont {D.}~\bibnamefont {Kirwan}},
  \bibinfo {author} {\bibfnamefont {A.}~\bibnamefont {Costa}}, \bibinfo
  {author} {\bibfnamefont {R.}~\bibnamefont {Muniz}}, \bibinfo {author}
  {\bibfnamefont {D.}~\bibnamefont {Mills}}, \ and\ \bibinfo {author}
  {\bibfnamefont {M.}~\bibnamefont {Ferreira}},\ }\href@noop {} {\bibfield
  {journal} {\bibinfo  {journal} {Physical Review B}\ }\textbf {\bibinfo
  {volume} {81}},\ \bibinfo {pages} {153408} (\bibinfo {year}
  {2010})}\BibitemShut {NoStop}%
\bibitem [{\citenamefont {Kr{\'a}l}\ and\ \citenamefont
  {Sadeghpour}(2002)}]{kral2002laser}%
  \BibitemOpen
  \bibfield  {author} {\bibinfo {author} {\bibfnamefont {P.}~\bibnamefont
  {Kr{\'a}l}}\ and\ \bibinfo {author} {\bibfnamefont {H.}~\bibnamefont
  {Sadeghpour}},\ }\href@noop {} {\bibfield  {journal} {\bibinfo  {journal}
  {Physical Review B}\ }\textbf {\bibinfo {volume} {65}},\ \bibinfo {pages}
  {161401} (\bibinfo {year} {2002})}\BibitemShut {NoStop}%
\bibitem [{\citenamefont {Brand{\~a}o}\ \emph {et~al.}(2015)\citenamefont
  {Brand{\~a}o}, \citenamefont {Moraes}, \citenamefont {Cunha}, \citenamefont
  {Lima},\ and\ \citenamefont {Filgueiras}}]{brandao2015inertial}%
  \BibitemOpen
  \bibfield  {author} {\bibinfo {author} {\bibfnamefont {J.~E.}\ \bibnamefont
  {Brand{\~a}o}}, \bibinfo {author} {\bibfnamefont {F.}~\bibnamefont {Moraes}},
  \bibinfo {author} {\bibfnamefont {M.}~\bibnamefont {Cunha}}, \bibinfo
  {author} {\bibfnamefont {J.~R.}\ \bibnamefont {Lima}}, \ and\ \bibinfo
  {author} {\bibfnamefont {C.}~\bibnamefont {Filgueiras}},\ }\href@noop {}
  {\bibfield  {journal} {\bibinfo  {journal} {Results in Physics}\ }\textbf
  {\bibinfo {volume} {5}},\ \bibinfo {pages} {55} (\bibinfo {year}
  {2015})}\BibitemShut {NoStop}%
\bibitem [{\citenamefont {Poncharal}\ \emph {et~al.}(2002)\citenamefont
  {Poncharal}, \citenamefont {Berger}, \citenamefont {Yi}, \citenamefont
  {Wang},\ and\ \citenamefont {de~Heer}}]{poncharal2002roomACS}%
  \BibitemOpen
  \bibfield  {author} {\bibinfo {author} {\bibfnamefont {P.}~\bibnamefont
  {Poncharal}}, \bibinfo {author} {\bibfnamefont {C.}~\bibnamefont {Berger}},
  \bibinfo {author} {\bibfnamefont {Y.}~\bibnamefont {Yi}}, \bibinfo {author}
  {\bibfnamefont {Z.~L.}\ \bibnamefont {Wang}}, \ and\ \bibinfo {author}
  {\bibfnamefont {W.~A.}\ \bibnamefont {de~Heer}},\ }\href {\doibase
  10.1021/jp021271u} {\bibfield  {journal} {\bibinfo  {journal} {The Journal of
  Physical Chemistry B}\ }\textbf {\bibinfo {volume} {106}},\ \bibinfo {pages}
  {12104} (\bibinfo {year} {2002})}\BibitemShut {NoStop}%
\bibitem [{\citenamefont {White}\ and\ \citenamefont
  {Todorov}(1998)}]{white1998carbon}%
  \BibitemOpen
  \bibfield  {author} {\bibinfo {author} {\bibfnamefont {C.~T.}\ \bibnamefont
  {White}}\ and\ \bibinfo {author} {\bibfnamefont {T.~N.}\ \bibnamefont
  {Todorov}},\ }\href@noop {} {\bibfield  {journal} {\bibinfo  {journal}
  {Nature}\ }\textbf {\bibinfo {volume} {393}},\ \bibinfo {pages} {240}
  (\bibinfo {year} {1998})}\BibitemShut {NoStop}%
\bibitem [{\citenamefont {Dresselhaus}\ \emph {et~al.}(1998)\citenamefont
  {Dresselhaus}, \citenamefont {Riichiro} \emph
  {et~al.}}]{dresselhaus1998physical}%
  \BibitemOpen
  \bibfield  {author} {\bibinfo {author} {\bibfnamefont {G.}~\bibnamefont
  {Dresselhaus}}, \bibinfo {author} {\bibfnamefont {S.}~\bibnamefont
  {Riichiro}},  \emph {et~al.},\ }\href@noop {} {\emph {\bibinfo {title}
  {Physical properties of carbon nanotubes}}}\ (\bibinfo  {publisher} {World
  scientific},\ \bibinfo {year} {1998})\BibitemShut {NoStop}%
\bibitem [{\citenamefont {Ando}(2000)}]{ando2000theory}%
  \BibitemOpen
  \bibfield  {author} {\bibinfo {author} {\bibfnamefont {T.}~\bibnamefont
  {Ando}},\ }\href@noop {} {\bibfield  {journal} {\bibinfo  {journal}
  {Semiconductor science and technology}\ }\textbf {\bibinfo {volume} {15}},\
  \bibinfo {pages} {R13} (\bibinfo {year} {2000})}\BibitemShut {NoStop}%
\bibitem [{\citenamefont {Lima}\ \emph {et~al.}(2014)\citenamefont {Lima},
  \citenamefont {Brand{\~a}o}, \citenamefont {Cunha},\ and\ \citenamefont
  {Moraes}}]{lima2014effects}%
  \BibitemOpen
  \bibfield  {author} {\bibinfo {author} {\bibfnamefont {J.~R.}\ \bibnamefont
  {Lima}}, \bibinfo {author} {\bibfnamefont {J.}~\bibnamefont {Brand{\~a}o}},
  \bibinfo {author} {\bibfnamefont {M.~M.}\ \bibnamefont {Cunha}}, \ and\
  \bibinfo {author} {\bibfnamefont {F.}~\bibnamefont {Moraes}},\ }\href@noop {}
  {\bibfield  {journal} {\bibinfo  {journal} {The European Physical Journal D}\
  }\textbf {\bibinfo {volume} {68}},\ \bibinfo {pages} {94} (\bibinfo {year}
  {2014})}\BibitemShut {NoStop}%
\bibitem [{\citenamefont {Lima}\ and\ \citenamefont
  {Moraes}(2015)}]{lima2015combined}%
  \BibitemOpen
  \bibfield  {author} {\bibinfo {author} {\bibfnamefont {J.~R.}\ \bibnamefont
  {Lima}}\ and\ \bibinfo {author} {\bibfnamefont {F.}~\bibnamefont {Moraes}},\
  }\href@noop {} {\bibfield  {journal} {\bibinfo  {journal} {The European
  Physical Journal B}\ }\textbf {\bibinfo {volume} {88}},\ \bibinfo {pages}
  {63} (\bibinfo {year} {2015})}\BibitemShut {NoStop}%
\bibitem [{\citenamefont {Cunha}\ \emph {et~al.}(2015)\citenamefont {Cunha},
  \citenamefont {Brand{\~a}o}, \citenamefont {Lima},\ and\ \citenamefont
  {Moraes}}]{cunha2015spin}%
  \BibitemOpen
  \bibfield  {author} {\bibinfo {author} {\bibfnamefont {M.~M.}\ \bibnamefont
  {Cunha}}, \bibinfo {author} {\bibfnamefont {J.}~\bibnamefont {Brand{\~a}o}},
  \bibinfo {author} {\bibfnamefont {J.~R.}\ \bibnamefont {Lima}}, \ and\
  \bibinfo {author} {\bibfnamefont {F.}~\bibnamefont {Moraes}},\ }\href@noop {}
  {\bibfield  {journal} {\bibinfo  {journal} {The European Physical Journal B}\
  }\textbf {\bibinfo {volume} {88}},\ \bibinfo {pages} {288} (\bibinfo {year}
  {2015})}\BibitemShut {NoStop}%
\bibitem [{\citenamefont {Hamada}\ \emph {et~al.}(2015)\citenamefont {Hamada},
  \citenamefont {Yokoyama},\ and\ \citenamefont {Murakami}}]{hamada2015spin}%
  \BibitemOpen
  \bibfield  {author} {\bibinfo {author} {\bibfnamefont {M.}~\bibnamefont
  {Hamada}}, \bibinfo {author} {\bibfnamefont {T.}~\bibnamefont {Yokoyama}}, \
  and\ \bibinfo {author} {\bibfnamefont {S.}~\bibnamefont {Murakami}},\
  }\href@noop {} {\bibfield  {journal} {\bibinfo  {journal} {Physical Review
  B}\ }\textbf {\bibinfo {volume} {92}},\ \bibinfo {pages} {060409} (\bibinfo
  {year} {2015})}\BibitemShut {NoStop}%
\bibitem [{\citenamefont {Matsuo}\ \emph {et~al.}(2011)\citenamefont {Matsuo},
  \citenamefont {Ieda}, \citenamefont {Saitoh},\ and\ \citenamefont
  {Maekawa}}]{matsuo2011spin}%
  \BibitemOpen
  \bibfield  {author} {\bibinfo {author} {\bibfnamefont {M.}~\bibnamefont
  {Matsuo}}, \bibinfo {author} {\bibfnamefont {J.}~\bibnamefont {Ieda}},
  \bibinfo {author} {\bibfnamefont {E.}~\bibnamefont {Saitoh}}, \ and\ \bibinfo
  {author} {\bibfnamefont {S.}~\bibnamefont {Maekawa}},\ }\href@noop {}
  {\bibfield  {journal} {\bibinfo  {journal} {Physical Review B}\ }\textbf
  {\bibinfo {volume} {84}},\ \bibinfo {pages} {104410} (\bibinfo {year}
  {2011})}\BibitemShut {NoStop}%
\bibitem [{\citenamefont {Da~Costa}(1981)}]{da1981quantum}%
  \BibitemOpen
  \bibfield  {author} {\bibinfo {author} {\bibfnamefont {R.}~\bibnamefont
  {Da~Costa}},\ }\href@noop {} {\bibfield  {journal} {\bibinfo  {journal}
  {Physical Review A}\ }\textbf {\bibinfo {volume} {23}},\ \bibinfo {pages}
  {1982} (\bibinfo {year} {1981})}\BibitemShut {NoStop}%
\bibitem [{\citenamefont {Santos}\ \emph {et~al.}(2016)\citenamefont {Santos},
  \citenamefont {Fumeron}, \citenamefont {Berche},\ and\ \citenamefont
  {Moraes}}]{Santos_2016}%
  \BibitemOpen
  \bibfield  {author} {\bibinfo {author} {\bibfnamefont {F.}~\bibnamefont
  {Santos}}, \bibinfo {author} {\bibfnamefont {S.}~\bibnamefont {Fumeron}},
  \bibinfo {author} {\bibfnamefont {B.}~\bibnamefont {Berche}}, \ and\ \bibinfo
  {author} {\bibfnamefont {F.}~\bibnamefont {Moraes}},\ }\href {\doibase
  10.1088/0957-4484/27/13/135302} {\bibfield  {journal} {\bibinfo  {journal}
  {Nanotechnology}\ }\textbf {\bibinfo {volume} {27}},\ \bibinfo {pages}
  {135302} (\bibinfo {year} {2016})}\BibitemShut {NoStop}%
\bibitem [{\citenamefont {Fumeron}\ \emph {et~al.}(2017)\citenamefont
  {Fumeron}, \citenamefont {Berche}, \citenamefont {Moraes},\ and\
  \citenamefont {Santos}}]{Fumeron_2017}%
  \BibitemOpen
  \bibfield  {author} {\bibinfo {author} {\bibfnamefont {S.}~\bibnamefont
  {Fumeron}}, \bibinfo {author} {\bibfnamefont {B.}~\bibnamefont {Berche}},
  \bibinfo {author} {\bibfnamefont {F.}~\bibnamefont {Moraes}}, \ and\ \bibinfo
  {author} {\bibfnamefont {F.}~\bibnamefont {Santos}},\ }\href {\doibase
  10.1088/1742-6596/785/1/012003} {\bibfield  {journal} {\bibinfo  {journal}
  {Journal of Physics: Conference Series}\ }\textbf {\bibinfo {volume} {785}},\
  \bibinfo {pages} {012003} (\bibinfo {year} {2017})}\BibitemShut {NoStop}%
\bibitem [{\citenamefont {Serafim}\ \emph {et~al.}(2019)\citenamefont
  {Serafim}, \citenamefont {Santos}, \citenamefont {Lima}, \citenamefont
  {Filgueiras},\ and\ \citenamefont {Moraes}}]{serafim2019position}%
  \BibitemOpen
  \bibfield  {author} {\bibinfo {author} {\bibfnamefont {F.}~\bibnamefont
  {Serafim}}, \bibinfo {author} {\bibfnamefont {F.}~\bibnamefont {Santos}},
  \bibinfo {author} {\bibfnamefont {J.~R.}\ \bibnamefont {Lima}}, \bibinfo
  {author} {\bibfnamefont {C.}~\bibnamefont {Filgueiras}}, \ and\ \bibinfo
  {author} {\bibfnamefont {F.}~\bibnamefont {Moraes}},\ }\href@noop {}
  {\bibfield  {journal} {\bibinfo  {journal} {Physica E: Low-dimensional
  Systems and Nanostructures}\ }\textbf {\bibinfo {volume} {108}},\ \bibinfo
  {pages} {139} (\bibinfo {year} {2019})}\BibitemShut {NoStop}%
\bibitem [{\citenamefont {Medina}\ \emph {et~al.}(2008)\citenamefont {Medina},
  \citenamefont {LÃ³pez},\ and\ \citenamefont {Berche}}]{0295-5075-83-4-47005}%
  \BibitemOpen
  \bibfield  {author} {\bibinfo {author} {\bibfnamefont {E.}~\bibnamefont
  {Medina}}, \bibinfo {author} {\bibfnamefont {A.}~\bibnamefont {L\'opez}}, \
  and\ \bibinfo {author} {\bibfnamefont {B.}~\bibnamefont {Berche}},\ }\href
  {http://stacks.iop.org/0295-5075/83/i=4/a=47005} {\bibfield  {journal}
  {\bibinfo  {journal} {EPL (Europhysics Letters)}\ }\textbf {\bibinfo {volume}
  {83}},\ \bibinfo {pages} {47005} (\bibinfo {year} {2008})}\BibitemShut
  {NoStop}%
\bibitem [{\citenamefont {Berche}\ \emph {et~al.}(2012)\citenamefont {Berche},
  \citenamefont {Medina},\ and\ \citenamefont {L\'opez}}]{0295-5075-97-6-67007}%
  \BibitemOpen
  \bibfield  {author} {\bibinfo {author} {\bibfnamefont {B.}~\bibnamefont
  {Berche}}, \bibinfo {author} {\bibfnamefont {E.}~\bibnamefont {Medina}}, \
  and\ \bibinfo {author} {\bibfnamefont {A.}~\bibnamefont {L\'opez}},\ }\href
  {http://stacks.iop.org/0295-5075/97/i=6/a=67007} {\bibfield  {journal}
  {\bibinfo  {journal} {EPL (Europhysics Letters)}\ }\textbf {\bibinfo {volume}
  {97}},\ \bibinfo {pages} {67007} (\bibinfo {year} {2012})}\BibitemShut
  {NoStop}%
\bibitem [{\citenamefont {Berche}\ \emph {et~al.}(2016)\citenamefont {Berche},
  \citenamefont {Malterre},\ and\ \citenamefont
  {Medina}}]{doi:10.1119/1.4955153}%
  \BibitemOpen
  \bibfield  {author} {\bibinfo {author} {\bibfnamefont {B.}~\bibnamefont
  {Berche}}, \bibinfo {author} {\bibfnamefont {D.}~\bibnamefont {Malterre}}, \
  and\ \bibinfo {author} {\bibfnamefont {E.}~\bibnamefont {Medina}},\ }\href
  {\doibase 10.1119/1.4955153} {\bibfield  {journal} {\bibinfo  {journal}
  {American Journal of Physics}\ }\textbf {\bibinfo {volume} {84}},\ \bibinfo
  {pages} {616} (\bibinfo {year} {2016})}\BibitemShut {NoStop}%
\bibitem [{\citenamefont {Berche}\ and\ \citenamefont
  {Medina}(2013)}]{0143-0807-34-1-161}%
  \BibitemOpen
  \bibfield  {author} {\bibinfo {author} {\bibfnamefont {B.}~\bibnamefont
  {Berche}}\ and\ \bibinfo {author} {\bibfnamefont {E.}~\bibnamefont
  {Medina}},\ }\href {http://stacks.iop.org/0143-0807/34/i=1/a=161} {\bibfield
  {journal} {\bibinfo  {journal} {European Journal of Physics}\ }\textbf
  {\bibinfo {volume} {34}},\ \bibinfo {pages} {161} (\bibinfo {year}
  {2013})}\BibitemShut {NoStop}%
\bibitem [{\citenamefont {Berche}\ \emph {et~al.}(2010)\citenamefont {Berche},
  \citenamefont {Chatelain},\ and\ \citenamefont
  {Medina}}]{0143-0807-31-5-026}%
  \BibitemOpen
  \bibfield  {author} {\bibinfo {author} {\bibfnamefont {B.}~\bibnamefont
  {Berche}}, \bibinfo {author} {\bibfnamefont {C.}~\bibnamefont {Chatelain}}, \
  and\ \bibinfo {author} {\bibfnamefont {E.}~\bibnamefont {Medina}},\ }\href
  {http://stacks.iop.org/0143-0807/31/i=5/a=026} {\bibfield  {journal}
  {\bibinfo  {journal} {European Journal of Physics}\ }\textbf {\bibinfo
  {volume} {31}},\ \bibinfo {pages} {1267} (\bibinfo {year}
  {2010})}\BibitemShut {NoStop}%
\bibitem [{\citenamefont {Hodge}\ \emph {et~al.}(2014)\citenamefont {Hodge},
  \citenamefont {Migirditch},\ and\ \citenamefont
  {Kerr}}]{doi:10.1119/1.4868094}%
  \BibitemOpen
  \bibfield  {author} {\bibinfo {author} {\bibfnamefont {W.~B.}\ \bibnamefont
  {Hodge}}, \bibinfo {author} {\bibfnamefont {S.~V.}\ \bibnamefont
  {Migirditch}}, \ and\ \bibinfo {author} {\bibfnamefont {W.~C.}\ \bibnamefont
  {Kerr}},\ }\href {\doibase 10.1119/1.4868094} {\bibfield  {journal} {\bibinfo
   {journal} {American Journal of Physics}\ }\textbf {\bibinfo {volume} {82}},\
  \bibinfo {pages} {681} (\bibinfo {year} {2014})}\BibitemShut {NoStop}%
\bibitem [{\citenamefont {Phillips}(2012)}]{Phillips}%
  \BibitemOpen
  \bibfield  {author} {\bibinfo {author} {\bibfnamefont {P.}~\bibnamefont
  {Phillips}},\ }\href@noop {} {\emph {\bibinfo {title} {Advanced Solid State
  Physics}}}\ (\bibinfo  {publisher} {Cambridge University Press},\ \bibinfo
  {year} {2012})\BibitemShut {NoStop}%
\bibitem [{\citenamefont {Newson}\ \emph {et~al.}(2011)\citenamefont {Newson},
  \citenamefont {Green}, \citenamefont {Hersam},\ and\ \citenamefont
  {Van~Driel}}]{newson2011coherent}%
  \BibitemOpen
  \bibfield  {author} {\bibinfo {author} {\bibfnamefont {R.}~\bibnamefont
  {Newson}}, \bibinfo {author} {\bibfnamefont {A.}~\bibnamefont {Green}},
  \bibinfo {author} {\bibfnamefont {M.~C.}\ \bibnamefont {Hersam}}, \ and\
  \bibinfo {author} {\bibfnamefont {H.}~\bibnamefont {Van~Driel}},\ }\href@noop
  {} {\bibfield  {journal} {\bibinfo  {journal} {Physical Review B}\ }\textbf
  {\bibinfo {volume} {83}},\ \bibinfo {pages} {115421} (\bibinfo {year}
  {2011})}\BibitemShut {NoStop}%
\end{thebibliography}

\appendix*
\section{Azimuthal magnetic field}
For the sake of completeness, we present here the results concerning an azimuthal magnetic field
\begin{equation}
\vec{B}=B\hat{\varphi} \label{Bazim}
\end{equation}
with $|\vec B|$ constant.
Obviously, this is a much more difficult experimental condition but hopefully it is much less interesting since there is no SO coupling.

\subsection{Energy}

The corresponding vector potential to (\ref{Bazim}) is given by $\vec{A}=-B\rho\hat{z}.$
The Hamiltonian can be written as
\begin{equation}
    \H=\frac{1}{2m}\Bigl[\left(-i\hbar a^{-1}\partial_{\varphi} \right)^2 +\left(-i\hbar \partial_z -|e|Ba \right)^2 -\frac{\hbar^2}{8ma^2}\Bigr] \boldsymbol{\sigma}_0+\frac{|e|\hbar}{2m} B \boldsymbol{\sigma}_{\varphi}-\omega a\left(-i\hbar a^{-1}\partial_{\varphi} \right)\boldsymbol{\sigma}_0-\frac{\hbar \omega}{2}\boldsymbol{\sigma}_z
    \label{h_azimuthal}
\end{equation}
where it has been taken into account the fact that due to Eq.~(\ref{eq-Eprime}), the spin-orbit interaction vanishes in the case of an azimuthal magnetic field.
Acting on a spinor (\ref{eq-spinor}), it yields  the same form as in the case of an axial magnetic field,
\begin{equation}
\H\Psi=\begin{pmatrix}
\hbar\Omega^- & -i\mu_BBe^{-i\varphi}\\
i\mu_BBe^{i\varphi}& 
\hbar\Omega^+
\end{pmatrix}\begin{pmatrix}\alpha e^{-i\varphi/2}\\ \beta  e^{i\varphi/2}\end{pmatrix} e^{i\ell\varphi}e^{ikz},
\end{equation}
 but now with the parametrization
\begin{eqnarray}
\hbar\Omega^\pm&=& 
\frac{\hbar^2}{2ma^2}(\ell\pm 1/2)^2+\frac 1{2m}(\hbar k-|e|Ba)^2-\hbar\omega\ell-\frac{\hbar^2}{8ma^2}, \label{EqA4}\\
\tan\theta&=&\frac{\omega_c}{\Omega^+-\Omega^-}
\label{EqA5}
\end{eqnarray}
where a cancellation occurs between the Zeeman and part of the orbital contributions as one can see by careful inspection. 
The parametrization being the same, the eigenvalues and eigenspinors are given by 
\begin{equation}
E_{\ell k\sigma}=\frac 12\hbar(\Omega^++\Omega^-)+\frac 12\sigma\hbar\sqrt{
(\Omega^+-\Omega^-)^2+4\mu_B^2B^2/\hbar^2
}.
\end{equation}
and 
Eqs.~(\ref{Eq-eigenstates}) with the appropriate modifications of the  $\Omega$'s. Note that the vector potential being now in the $z$ direction, the periodic repetition of parabolas in the energy spectrum would be obtained at various $k$-values rather than $\ell$-values.

\begin{figure}
	\centering
        \hspace{-1mm}\includegraphics[scale=1.2]{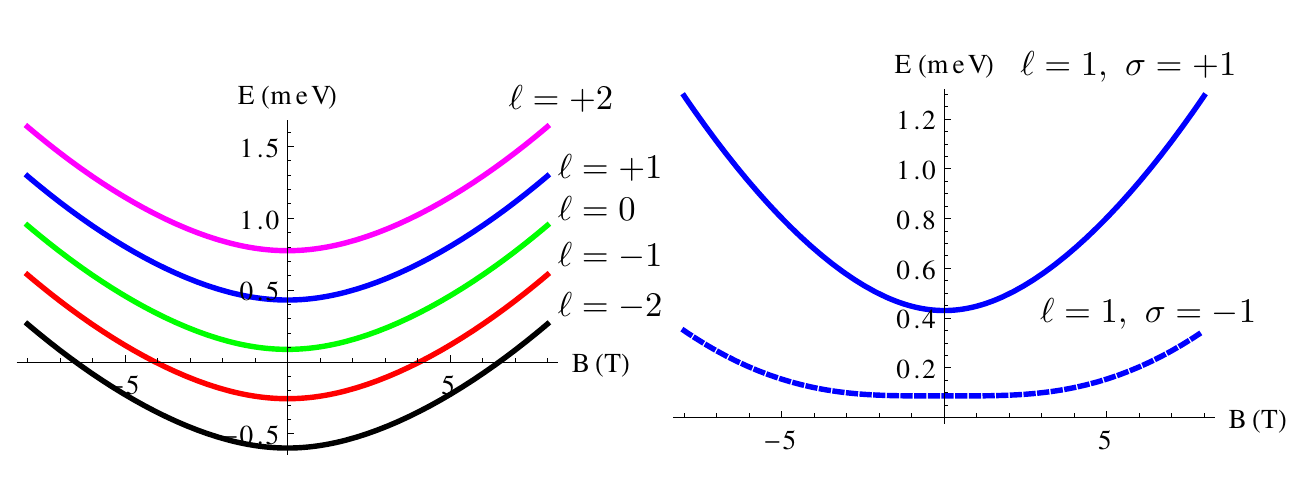}
		\hspace{-26mm}\includegraphics[scale=1.4]{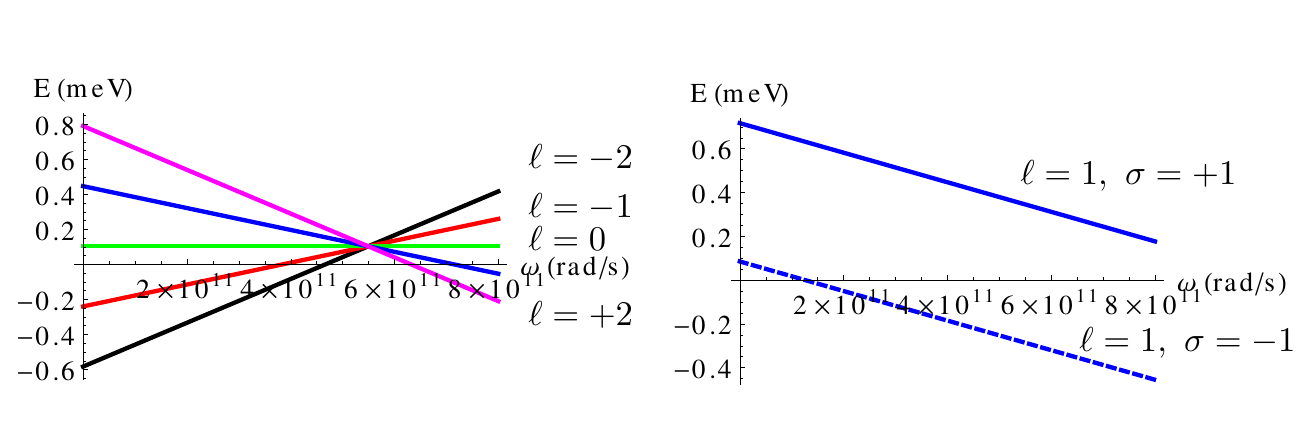}
\caption{Energy (in meV), top:  as function of the magnetic field, for the azimuthal magnetic field case, when $\omega=10^{10} $rad.s$^{-1}$, bottom: as function of the angular velocity, when $B=1 $T. The values of $\ell, k,\sigma$ are indicated as plot legends. The radius of the nanotube is fixed to $a=50$~nm, the right plots show for the level $\ell=1$, ($k=0$), the two values of $\sigma=-1,+1$. 
}\label{Fig2}
\end{figure}

 Using the same figures as in the previous section, we find of course similar orders of magnitude for various contributions to the energy.
Since there is no spin-orbit coupling in this case, the form of the term inside the square root is different, without the combined term involving rotation and field, as in the previous case.  Fig. \ref{Fig2} gives an idea of the energy behaviour for a few states.

\subsection{Charge currents} 
 In order to obtain the charge and spin currents, we again need  to obtain $ {\mathbf v}_{\varphi}$ and $ {\mathbf v}_z$.
From the Hamiltonian (\ref{h_azimuthal}), we can find that
\begin{equation}
     {\mathbf v}_{\varphi}=\frac{1}{m}\left(-i\hbar a^{-1}\partial_{\phi} \right)\boldsymbol{\sigma}_0-\omega a \boldsymbol{\sigma}_0,
\end{equation}
and
\begin{equation}
     {\mathbf v}_z=\frac{1}{m}\left(-i\hbar \partial_z -|e|Ba \right)\boldsymbol{\sigma}_0
\end{equation}
and  the charge currents follow. They are given by
\begin{eqnarray}
J_{\varphi,\ell k \sigma}&=&-|e|\Biggl[\frac{\hbar \ell}{ma}-\omega a -\sigma\frac\hbar{2ma}\cos\theta \Biggl],
\\
    J_{z,\ell k \sigma}&=&-|e|\Biggl[ \frac{\hbar k}m -\frac{|e|Ba}m \Biggl], \label{Jzazim}
\end{eqnarray}
where $\theta$ depends both on $\omega$ and $B$ (see Eqs. (\ref{EqA5}) and (\ref{EqA4})).
 It is interesting to note that, in the case of the azimuthal magnetic field, the  contributions from the   rotation appear solely in the $\varphi$-component of the charge current. Furthermore, the spin polarization $\sigma$ appears only in the $\varphi$-component. 
\subsection{Spin currents}
Now, we will write the expressions for the spin currents.
In the $\varphi$ direction, we have
\begin{equation}
{S}^z_{\varphi,\ell k \sigma}=\frac{\hbar}{2} \Biggl[\sigma
\Biggl(\frac{\hbar\ell}{ma}-\omega a\Biggr)\cos\theta
-\frac{\hbar}{2ma}
\Biggr].
\end{equation}
In the $z$ direction, we have
\begin{equation}
{S}^z_{z,\ell k \sigma}=\frac{\hbar}{2} \Biggl[\sigma
\Biggl(\frac{\hbar k}m -\frac{|e|Ba}{m}\Biggl)\cos\theta\Biggr] = \frac{\hbar\sigma}{2} \frac{J_{z,\ell k \sigma}}{(-|e|)}\cos\theta . \label{Szazim}
\end{equation}
Here, differently from the charge current, both spin current components depend on $\omega$, $B$ and the spin polarization $\sigma$. From Eqs. (\ref{Jzazim}) and (\ref{Szazim}) it is clear that, if ${J}^z_{z,\ell k \sigma}$ is tuned to zero  by adjusting $B$, the corresponding spin current component ${S}^z_{z,\ell k \sigma}=0$.

\end{document}